\documentclass[11pt,a4paper]{article} 
\usepackage{amsmath} 
\usepackage{amsthm} 
\usepackage{amssymb}	
\usepackage{graphicx} 
\usepackage{multicol} 
\usepackage[dvips,letterpaper,margin=1 in,bottom=0.5in]{geometry}
\usepackage{color}
\usepackage{setspace}
\usepackage{indentfirst} 
\usepackage{textcomp}
\usepackage{empheq}
\usepackage{breqn}
\usepackage{mathrsfs}
\usepackage[utf8]{inputenc}
\makeatletter 
\renewcommand{\maketitle} 
{ \begingroup \vskip 10pt \begin{center} \large {\bf \@title}
	\vskip 10pt \large \@author \hskip 20pt \@date \end{center}
  \vskip 10pt \endgroup \setcounter{footnote}{0} }
\makeatother 
 
\newcommand{\pd}{\partial} 

\newcommand{\be}{\begin{equation}}
\newcommand{\ee}{\end{equation}}
 
 \newcommand{\gld}[1]{\widehat{\mathcal{L}}_{#1}}

 \newcommand{\bkt}[1]{\left(#1\right)}
 \newcommand{\gm}{\mathcal{H}}
\newcommand{\sbkt}[1]{\left[#1\right]}
\let\baraccent=\= 
\renewcommand{\=}[1]{\stackrel{#1}{=}} 

\theoremstyle{definition}

\theoremstyle{remark}

	\numberwithin{equation}{section} 
	\numberwithin{figure}{section} 
	\numberwithin{table}{section} 
	
	\usepackage{jheppub} 
	
	\usepackage[T1]{fontenc} 

	\title{\boldmath Canonical formulation and conserved charges of double field theory}
	\author{Usman Naseer}
	\affiliation{Center for Theoretical Physics,\\
	Massachusetts Institute of Technology,\\
	Cambridge, MA 02139, USA.
	}
	\emailAdd{unaseer@mit.edu}
	
	\abstract{ We provide the  canonical formulation of double field theory. It is shown that this dynamics is subject to primary and secondary constraints. The Poisson bracket algebra of secondary constraints is  shown to close on-shell according to the C-bracket. A systematic way of writing boundary integrals in doubled geometry is given. By including appropriate boundary terms in the double field theory Hamiltonian, expressions for conserved energy and momentum of an asymptotically flat doubled space-time are obtained and applied to a number of solutions.
	
	}
	
\begin{document}
	\rightline{}
	\rightline{\tt }
	\rightline{\tt  MIT-CTP/4696}
    \rightline{August 2015}
	\onehalfspacing
	
	\maketitle
	\section{Introduction}\label{sec:intro}
    Double field theory was developed to make manifest the $O(d,d)$ T-duality  symmetry in the low energy effective field theory limit of string theory obtained after compactification on  a $d$-dimensional torus.  In addition to the usual space-time coordinates, `winding' coordinates are introduced. The metric and the Kalb-Ramond two-form are combined into a `generalized metric'. This generalized metric transforms linearly under global $O(d,d)$ transformations. Gauge transformations of the fields can also be written in an $O(d,d)$ covariant form and they can be interpreted as the `generalized coordinate transformations' in the doubled space-time. The action of double field theory, written in terms of the generalized metric, is then manifestly invariant under these transformations. Double field theory is a restricted theory. The so-called strong constraint restricts the theory to live on a $d$-dimensional subspace of the doubled space-time. Different solutions of the strong constraint are then related by T-duality.
    
    Double field theory was developed in \cite{siegel_superspace_1993,siegel_twovierbein_1993,hull_double_2009a,hohm_background_2010a,hull_gauge_2009a,hohm_generalized_2010a} and earlier ideas can be found in \cite{tseytlin_duality_1990,tseytlin_duality_1991,duff_duality_1990,duff_duality_1990a}. Further developments of double field theory are discussed in \cite{hohm_framelike_2011,kwak_invariances_2010a,hohm_tduality_2011,hohm_double_2011a,hohm_unification_2011,hohm_double_2011b,hohm_massive_2011a,hohm_n1_2012a,hohm_riemann_2012a,hohm_gauge_2013,hohm_large_2013a,hull_finite_2015a,naseer_note_2015a,berman_generalized_2011,berman_boundary_2011,berman_so55_2011,berman_local_2012,berman_global_2014,west_e11_2010,rocen_e11_2010,jeon_differential_2011a,jeon_stringy_2011,jeon_incorporation_2011a,jeon_supersymmetric_2012a,jeon_ramond_2012,jeon_comments_2013a,schulz_tfolds_2012,copland_connecting_2012,copland_double_2012,thompson_duality_2011a,albertsson_double_2011a,aldazabal_effective_2011a,geissbuhler_double_2011a,grana_gauged_2012a,coimbra_supergravity_2011,coimbra_edd_2011,vaisman_geometry_2012a}. For reviews on this subject see \cite{zwiebach_double_2012a,aldazabal_double_2013b,hohm_spacetime_2013a}.
    
    The most geometrical formulation of double field theory is in terms of the generalized metric and the dilaton. The action for double field theory, up to  boundary terms, on a $2d$-dimensional doubled space with generalized metric $\mathcal{H}_{MN}$ and dilaton $d$ can be written as:
    \begin{eqnarray}
    S_{\text{DFT}}=\int d^{2d}X\ e^{-2d}\mathcal{R}\bkt{d,\mathcal{H}_{MN}},
    \end{eqnarray}
    where $\mathcal{R}$ is the generalized scalar curvature which is a function of the dilaton and the generalized metric. By demanding no dependence on the dual coordinates, this action reduces to the low energy effective action of the NS-NS sector of bosonic string theory\cite{zwiebach_double_2012a}.
    
    Our aim in this paper is to provide the canonical formulation of double field theory. Due to many similarities with the ADM(Arnowitt-Deser-Misner) formulation of general relativity, it is instructive to briefly review this formulation before we discuss the canonical formulation for double field theory.
    
    In the case of general relativity, one starts with the Einstein-Hilbert action,
    \begin{eqnarray}S_{\text{EH}}=\int d^4x\sqrt{-g} R\sbkt{g_{\mu\nu}},\label{eq:introEH}\end{eqnarray}
    with $g_{\mu\nu}$ the metric on four dimensional space-time and $R\sbkt{g_{\mu\nu}}$ the Ricci scalar. The space-time manifold is foliated into space-like hyper-surfaces of constant time $t$, denoted by $\Sigma_t$. Space-time coordinates are split into time and space parts as:
    \begin{eqnarray}
x^{\mu}=\bkt{t,x^i}, \ \ \text{where}\ \ \ i=1,2,3.\label{eq:grcoordsplit}
\end{eqnarray}
One can now define a purely spatial metric $h_{ij}$ on the space-like hyper-surface by introducing the lapse function $\mathrm{N}$ and the shift vector $\mathcal{N}^{i}$\footnote{By a slight abuse of notation, we will use same symbols to denote the lapse function and shift vector for double field theory}. The metric on full space-time, $g_{\mu\nu}$, can then be expressed in terms of the spatial metric, the lapse function and the shift vector as follows:
\begin{eqnarray}
g_{00}=h_{ij}\mathcal{N}^i\mathcal{N}^j-\mathrm{N}^2, \ \ \ \ \ \ \ \ g_{0i}=g_{i0}=\mathcal{N}_i,\ \ \ \ \ \  g_{ij}=h_{ij}\ \ \ \ \text{and}\ \ \ \  \sqrt{-g}=\mathrm{N}\sqrt{h}.\label{eq:grmetsplit}
\end{eqnarray}
In terms of the variables ($\mathrm{N},\mathcal{N}^i,h_{ij}$), the action (\ref{eq:introEH}) takes the following form:
\begin{dmath}
S_{\text{EH}}=\int dtd^{3}x\ \sqrt{h}\bkt{\mathrm{N}^{-1}\frac{1}{4}\sbkt{\bkt{\pd_th_{ij}-\mathcal{L}_{\mathcal{N}}h_{ij}}\bkt{\pd_t h^{ij}-\mathcal{L}_{\mathcal{N}}h^{ij}}-h^{ij}h^{kl}\bkt{\pd_th_{ij}-\mathcal{L}_{\mathcal{N}}h_{ij}}\bkt{\pd_th_{kl}-\mathcal{L}_{\mathcal{N}}h_{kl}}}+\mathrm{N}\ ^{(3)}R},\label{eq:introEHsplit}
\end{dmath}
where $ ^{(3)}R$ denotes the Ricci scalar on the space-like hyper-surface and $\mathcal{L}_{\mathcal{N}}$ is the Lie derivative with respect to the shift vector $\mathcal{N}^{i}$. 

From the action (\ref{eq:introEHsplit}), one can compute canonical momenta and perform a Legendre transform to compute the Hamiltonian. The key aspect of this procedure is the emergence of primary and secondary constraints. The canonical momenta conjugate to the lapse function and the shift vector vanish identically. These are the primary constraints arising in the ADM formalism. The consistency of primary constraints, i.e., their invariance under time evolution, leads to two secondary constraints, $B\bkt{x}=0$ and $C_{i}\bkt{x}=0$. Precise expressions for these constraints can be found in \cite{arnowitt_dynamics_2008,blau_lectures_2014}. The Hamiltonian for general relativity can then be expressed in terms of these secondary constraints:
\begin{eqnarray}
H=\int d^{3}x\ \bkt{\mathrm{B}\sbkt{\mathrm{N}}+\mathrm{C}\sbkt{\mathcal{N}}}, \label{eq:grHam}
\end{eqnarray}
where $\mathrm{B}\sbkt{\lambda}$ and $\mathrm{C}\sbkt{\beta}$ are the so called `smeared' constraints built out of the `bare' secondary constraints as follows:
\begin{eqnarray}
\mathrm{B}\sbkt{\lambda}\equiv\int d^3x\ \lambda B,\ \ \ \text{and}\ \ \ \mathrm{C}\sbkt{\beta}\equiv\int d^3x\ \beta^{i}C_{i}.
\end{eqnarray}

It can be shown that the invariance of secondary constraints under time evolution is equivalent to the on-shell closure of the Poisson bracket algebra of the `smeared' constraints. This algebra is computed in great detail in \cite{dirac_lectures_2001} and it is :
\begin{eqnarray}
\{\mathrm{C}\sbkt{\beta_1},\mathrm{C}\sbkt{\beta_2}\}&=&\mathrm{C}\sbkt{\sbkt{\beta_1,\beta_2}}\\
\{\mathrm{B}\sbkt{\lambda},\mathrm{B}\sbkt{\rho}\}&=&\mathrm{C}\sbkt{\gamma},\ \ \ \text{where}\ \ \ \gamma^{i}\equiv h^{ij}\bkt{\lambda\pd_j\rho-\rho\pd_j\lambda}\\
\{\mathrm{B}\sbkt{\lambda},\mathrm{C}\sbkt{\beta}\}&=&\mathrm{B}\sbkt{-\beta^i\pd_i\lambda}\label{eq:calgGR},
\end{eqnarray}
where $\sbkt{\ ,\ }$ denotes the Lie bracket of vector fields.
    
    Since any solution of general relativity has to satisfy these primary and secondary constraints, we conclude that the on-shell value of the Hamiltonian (\ref{eq:grHam}) is zero. However, if the space-like hyper-surface has a non trivial boundary $\pd\Sigma$, one has to add appropriate boundary terms to the Hamiltonian. These boundary terms can lead to a non-zero value of the on-shell Hamiltonian \cite{Regge1974286}:
    \begin{eqnarray}
H_{\text{on-shell}}=H_{\text{bdy}}=\int d^2x\sqrt{\sigma}\sbkt{\mathrm{N}\bkt{h^{kj}n^i\pd_ih_{kj}-h^{ij}n^k\pd_ih_{kj}}+2\mathcal{N}_in_j\pi^{ij}},
\end{eqnarray}
where $\pi^{ij}$ is the canonical momentum conjugate to the metric $h_{ij}$ that can be computed easily from (\ref{eq:introEHsplit}). $\sigma_{\bar{i}\bar{j}}$, $\bar{i}=1,2$ is the metric on the two dimensional boundary $\pd\Sigma$ of the space-like hyper-surface. $n^i$ is the unit normal vector to the boundary.

Now, one can define notions of conserved charges, in particular the ADM energy and momentum. In general relativity, with vanishing cosmological constant, ADM energy and momentum are defined for space-times which are asymptotically flat. This is done by identifying a parameter $r$ in the metric such that the curved metric reduces to Minkowski metric when the parameter $r$ approaches $\infty$. In general $r$ is function of space-time coordinates. ADM energy is then obtained by setting the shift vector  to zero in the on-shell Hamiltonian and taking the limit $r\rightarrow \infty$. ADM momentum is obtained by setting the lapse function  to zero. The ADM energy and momentum can then be identified as  conserved charges associated with time and space translations respectively at $r\to \infty$. The ADM energy and momentum are given by:
\begin{eqnarray}
E_{\text{ADM}}&=&\lim_{r\to\infty}\int d^2x\sqrt{\sigma}\bkt{h^{kj}n^i\pd_ih_{kj}-h^{ij}n^k\pd_ih_{kj}},\\
P_{\text{ADM}}^i&=&2\lim_{r\to\infty}\int d^2x\sqrt{\sigma} n_j\pi^{ij}.
\end{eqnarray}
This completes our quick review of the ADM formulation of general relativity.  We have glossed over a lot of the details and intricacies of this formalism. In particular the notion of asymptotic flatness needs to be handled very carefully. We refer the interested reader to \cite{blau_lectures_2014,Regge1974286,arnowitt_dynamics_2008,dirac_lectures_2001} for a detailed exposition of the ADM formalism and related concepts.

    For the case of double field theory, a similar story unfolds. Our starting point is the double field theory action on a $2D$-dimensional doubled space, with generalized metric $\widehat{\gm}_{\hat{M}\hat{N}}$ and the dilaton $\widehat{d}$, where $\hat{M}=1,2,\cdots,2D$. We split the coordinates on the $2D$-dimensional manifold into temporal and spatial parts as: 
    \begin{eqnarray}
    X^{\hat{M}}=\bkt{\tilde{t},t,X^M},
    \end{eqnarray}
    and demand that fields and parameters are independent of the dual time coordinate $\tilde{t}$. $X^M$ are coordinates on the $2d$-dimensional doubled hyper-surface ($d=D-1$). This split allows an ADM-like decomposition of the full generalized metric $\widehat{\gm}_{\hat{M}\hat{N}}$ into the following:
    \begin{itemize}
    \item $\mathcal{\gm}_{MN}$, the induced generalized metric on the $2d$-dimensional doubled hyper-surface.
    \item  $\mathcal{N}^{M}$, the generalized shift vector.
    \item $\mathrm{N}$, the generalized lapse function which behaves as scalar under generalized diffeomorphisms of the doubled hyper-surface.
    \end{itemize}
    The dilaton $\widehat{d}$ of the $2D$-dimensional manifold is redefined as $e^{-2\widehat{d}}=\mathrm{N}e^{-2d}$ such that $d$ behaves as a density with respect to diffeomorphisms of the $2d$-dimensional hyper-surface. In terms of this new set of dynamical variables ($\mathrm{N},\mathcal{N}^M,d,\gm_{MN}$), the action of the double field theory reads:
\begin{dmath}
S=\int dt\int d^{2d}X \bkt{-\mathrm{N}^{-1}e^{-2d}\bkt{4\bkt{\mathcal{D}_td}^2+\frac{1}{8}\mathcal{D}_t\gm_{MN}\mathcal{D}_t\gm^{MN}}+\mathrm{N}e^{-2d}\mathcal{R}\bkt{d,\gm_{MN}}},\label{eq:Sintro}
\end{dmath}
where $\mathcal{D}_{t}\equiv \pd_t-\gld{\mathcal{N}}$ with $\gld{\mathcal{N}}$ denoting the generalized lie derivative with respect to $\mathcal{N}^M$. Similarities between this action and the one for general relativity  given in equation (\ref{eq:introEHsplit}) are obvious. Indeed one can show that (\ref{eq:Sintro}) reduces to (\ref{eq:introEHsplit}) upon proper truncation.

Following the usual procedure of computing the canonical momenta and doing the Legendre transform, we can obtain the Hamiltonian for double field theory. In that process, we find that the canonical momenta conjugate to the lapse function and the shift vector (denoted by $\Pi_{\mathrm{N}}$ and $\Pi^M$ respectively) vanish. This puts constraints on the dynamical variables of the theory, called primary constraints. The invariance of these constraints under time evolution leads to secondary constraints, $\mathscr{B}\bkt{X}=0$ and $\mathscr{C}_{M}\bkt{X}=0$. The precise form of these constraints in terms of the dynamical variables is given in equations (\ref{eq:defB}) and (\ref{eq:Const2Sol}) respectively. We introduce the convenient notion of `smeared' constraints by integrating $\mathscr{B}\bkt{X}$ and $\mathscr{C}_{M}\bkt{X}$ against suitable test functions $\lambda$ and $\xi^M$, as follows:
\begin{eqnarray}
\mathbf{B}\sbkt{\lambda}\equiv\int d^{2d}X\ \lambda\ \mathscr{B},\ \ \ \ \mathbf{C}\sbkt{\xi}\equiv\int d^{2d}X\ \xi^M\ \mathscr{C}_M,
\end{eqnarray}
where $\lambda$ and $\xi^M$ are smooth functions of coordinates such that the above integrals are well defined. The Hamiltonian of the double field theory, up to boundary terms, can then be written as:
\begin{dmath}
\mathbf{H}=\mathbf{B}\sbkt{\mathrm{N}}+\mathbf{C}\sbkt{\mathcal{N}}.\label{eq:introHam}
\end{dmath}

Consistency of the theory requires that the secondary constraints are also preserved under time evolution. The invariance of secondary constraints under time evolution is equivalent to the closure of the Poisson bracket algebra of the smeared constraints, on-shell. This algebra is computed in detail in section \ref{sec:algebra} and it is shown that it closes on-shell, as required. The constraint algebra is given by:
\begin{eqnarray}
\{\mathbf{C}\sbkt{\xi_1},\mathbf{C}\sbkt{\xi_2}\}&=&\mathbf{C}\sbkt{\sbkt{\xi_1,\xi_2}_C},\\
\{\mathbf{B}\sbkt{\lambda},\mathbf{B}\sbkt{\rho}\}&=&\mathbf{C}\sbkt{\chi},\ \ \ \text{where}\ \ \ \chi^M\equiv \gm^{MN}\bkt{\lambda\pd_N\rho-\rho\pd_N\lambda},\\
\{\mathbf{B}\sbkt{\lambda},\mathbf{C}\sbkt{\xi}\}&=&\mathbf{B}\sbkt{-\xi^P\pd_P\lambda},
\end{eqnarray}
 where $\sbkt{\ ,\ }_C$ denotes the C-bracket  defined as:
\begin{eqnarray}
\sbkt{\xi_1,\xi_2}_{\text{C}}^{M}=\xi_{1}^P\pd_P\xi_2^M-\frac{1}{2}\xi_{1P}\pd^M\xi_2^P-\bkt{1\leftrightarrow 2}.\label{eq:defCbra}
\end{eqnarray}
Again we see the similarities between the constraint algebra of double field theory and that of general relativity. In particular the Lie bracket and the metric $h_{ij}$ on space-like hyper-surface  are replaced by the generalized Lie bracket and the generalized metric $\gm_{MN}$ on the doubled hyper-surface as expected.
The Hamiltonian in equation (\ref{eq:introHam}) does not contain boundary terms. Since any solution of the double field theory has to satisfy primary and secondary constraint, the bulk Hamiltonian of equation (\ref{eq:introHam}) vanishes on-shell. However, if the $2d$-dimensional doubled hyper-surface has a non trivial boundary then the expression for Hamiltonian needs to be modified by adding appropriate boundary terms. The full Hamiltonian of the double field theory is then, $\mathbf{H}_{\text{DFT}}=\mathbf{H}+\mathbf{H}_{bdy}$. The importance of the boundary terms is evident because they give the on-shell value of the Hamiltonian. The boundary Hamiltonian is given in equation (\ref{eq:Hbdy}).

Motivated by the constructions of ADM energy and momenta in general relativity, we introduce conserved energy and momenta in double field theory. These conserved quantities are defined for doubled space-times which are asymptotically flat. To make precise the notion of flatness for a doubled space-time, we assume that the full generalized metric $\widehat{\gm}_{\hat{M}\hat{N}}$ depends on a function of coordinates $\mathscr{P}$ in such a way that it assumes the flat form $\widehat{\delta}_{\hat{M}\hat{N}}$ in the limit $\mathscr{P}\to\infty$. $\widehat{\delta}_{\hat{M}\hat{N}}$ is the Minkowski-type metric of signature $(2,2d)$. Let $X^{M}$, $M=1,\cdots,2d$ be the coordinates on the $2d$-dimensional doubled hyper-surface and $Y^{\bar{M}}$, $\bar{M}=1,\cdots,2d-1$ be the coordinates on its boundary. Due to the strong constraint, for a particular solution of double field theory, fields can only depend on a $d$-dimensional sub-space $\mathcal{M}_1$ of the $2d$-dimensional doubled hyper-surface $\mathcal{M}$. Also, if the boundary of $\mathcal{M}$ is characterized by the constraint $\mathscr{S}\bkt{X}=\text{constant}$, then  it can be shown that $\mathscr{S}$ can only depend on the `allowed' sub-space $\mathcal{M}_1$.  With these considerations in mind, expressions for the conserved energy and the conserved momentum take the following form:
\begin{eqnarray}
\mathbf{E}&=& \lim_{\mathscr{P}\to\infty} \int_{\pd\mathcal{M}_1} d^{d-1}Y\ \left|\frac{\pd X}{\pd X^{\prime}}\right|e^{-2d} \mathbf{N}_L\bkt{4\gm^{LP}\pd_Pd-\pd_{P}\gm^{LP}},\\
\mathbf{P}_M&=&\lim_{\mathscr{P}\to\infty}
\int_{\pd\mathcal{M}_1} d^{d-1}Y\ \left|\frac{\pd X}{\pd X^{\prime}}\right|\sbkt{2\gm_{MK}\Pi^{KL}\mathbf{N}_L-\frac{1}{4}\mathbf{N}_M\Pi_{d}},
\end{eqnarray}
where $\mathbf{N}_M$ is the gradient vector which characterizes the boundary and it is equal to $\pd_M\mathscr{S}\bkt{X}$. $X^{\prime M}$ are the coordinates adapted to the boundary, i.e.,
\begin{eqnarray}
X^{\prime M}=\bkt{Y^{\bar{M}},\mathscr{S}}.
\end{eqnarray}
Properties of gradient vectors are discussed in great detail in the appendix \ref{App:GenStoke}.

Using the expressions obtained for the conserved energy and momentum, one can compute conserved charges associated with specific solutions of double field theory. We apply our formulae to compute conserved charges for double field theory monopole and generalized pp-wave solutions discussed in  \cite{berkeley_strings_2014,berman_branes_2015}, and confirm the physical interpretation given there for various free parameters.

This paper is organized as follows. In section \ref{sec:dftnsplit} we briefly review the formulation of double field theory in terms of the generalized metric and the frame field and obtain the action (\ref{eq:Sintro}) by splitting the space-time of double field theory into temporal and spatial parts. In section \ref{sec:CanForm} we present the canonical formulation of double field theory and compute the bulk Hamiltonian. Section \ref{sec:algebra} is devoted to the discussion of constraints arising in the canonical formulation. We compute the algebra of secondary constraints under Poisson brackets and show that it closes on-shell. In section \ref{sec:conserved} we compute the boundary contribution to the double field theory Hamiltonian and define conserved charges. Conserved charges for some known solutions of double field theory are also computed there. Finally we conclude and summarize our results in section \ref{sec:conclusions}
	\section{Double field theory and its space/time split}\label{sec:dftnsplit}
   In this section we review important facts about double field theory and re-write its action in a form better suited for our later computations. We start by reviewing the formulation of double field theory in terms of the generalized metric. Afterwards we briefly review the formulation of double field theory in terms of the frame field. Finally we discuss how to split the space-time of double field theory explicitly into spatial and temporal parts. This split is accompanied by an ADM-like decomposition of the generalized metric and a re-definition of the dilaton. 
   
    \subsection{Generalized metric formulation}\label{subsec:genmet}
    Double field theory is an effective description of the massless bosonic sector of closed string theory which makes the T-duality symmetry manifest. It does so by introducing an additional set of $d$ coordinates, $\tilde{x}$, conjugate to the winding modes of the string. The total of $2d$ coordinates are combined into the $2d$-dimensional generalized coordinate vector $X^{M}$ as follows.
    \begin{eqnarray}
X^M=\begin{pmatrix}
\tilde{x}_i\\ x^i\label{eq:coord}
\end{pmatrix}.
\end{eqnarray}
The index $M$ is raised and lowered with the $O\bkt{d,d}$ invariant metric and its inverse defined as
\begin{eqnarray}
\eta_{MN}=\begin{pmatrix}
0 &\delta^i_{\ j}\\
\delta_i^{\ j}&0
\end{pmatrix},\ \ \ \ \ \ \ 
\eta^{MN}=\begin{pmatrix}
0 & \delta_i^{\ j} \\ \delta^{i}_{\ j}&0
\end{pmatrix}\label{eq:eta}.
\end{eqnarray}
Now the action of double field theory can be written in terms of the generalized metric \cite{hohm_generalized_2010a} as
\begin{eqnarray}
S_{\text{DFT}}&=&\int d^{2d}X \ \mathscr{L}_{\text{DFT}}\bkt{d,\gm_{MN}},\label{eq:Sdft}
\end{eqnarray}
where
\begin{dmath}
\mathscr{L}_{\text{DFT}}\bkt{d,\gm_{MN}}= e^{-2d}  \left(\frac{1}{8}\mathcal{H}^{MN}\partial_M\mathcal{H}^{KL}\partial_N\mathcal{H}_{KL}-\frac{1}{2}\gm^{MN}\pd_N\gm^{KL}\pd_L\gm_{MK} +4\gm^{MN}\pd_M d\pd_Nd-2\pd_M\mathcal{H}^{MN}\pd_Nd\right).\label{eq:Ldft}
\end{dmath}
$\gm_{MN}$ is called the generalized metric which combines the usual space-time metric $g_{ij}$ and the Kalb-Ramond field $b_{ij}$ into a symmetric $O\bkt{d,d}$ tensor given by:
\begin{eqnarray}
\gm_{MN}=\begin{pmatrix}
g^{ij}& -g^{ik}b_{kj}\\
b_{ik}g^{kj}&g_{ij}-b_{ik}g^{kl}b_{lj}
\end{pmatrix},\label{eq:genMet}
\end{eqnarray}
and it satisfies following constraints.
\begin{eqnarray}
\gm_{MP}\gm^{PN}=\delta_{M}^{\ N},\ \ \ \ \ \ \gm_{MP}\eta^{PQ}\gm_{QN}=\eta_{MN}.\label{eq:GenMetProp}
\end{eqnarray}
The dilaton $d$ is related to the scalar dilaton $\phi$ of the effective action by:
\begin{eqnarray}
e^{-2d}=\sqrt{-g}e^{-2\phi}.\label{eq:dilatonDef}
\end{eqnarray}
The action (\ref{eq:Sdft}) can also be expressed in terms of the generalized scalar curvature $\mathcal{R}\bkt{d,\gm_{MN}}$ up to some boundary terms. Indeed the Lagrangian density (\ref{eq:Ldft}) and the generalized scalar curvature are related as in
\begin{eqnarray}
\mathscr{L}_{\text{DFT}}\bkt{d,\gm_{MN}}=e^{-2d}\mathcal{R}\bkt{d,\mathcal{H}_{MN}}+\pd_{M}\bkt{e^{-2d}\sbkt{\pd_N\mathcal{H}^{MN}-4\gm^{MN}\pd_Nd}}.\label{eq:L+bterms}
\end{eqnarray}
The generalized curvature scalar can be obtained from this relation and it reads,
\begin{dmath}
\mathcal{R}\bkt{d,\mathcal{H}_{MN}}=\frac{1}{8}\mathcal{H}^{MN}\partial_M\mathcal{H}^{KL}\partial_N\mathcal{H}_{KL}-\frac{1}{2}\gm^{MN}\pd_N\gm^{KL}\pd_L\gm_{MK} +4\gm^{MN}\pd_M\pd_Nd+4\pd_M\mathcal{H}^{MN}\pd_Nd-4\gm^{MN}\pd_Md\pd_Nd-\pd_M\pd_N\gm^{MN}.\label{eq:GenRic}
\end{dmath}

Double field theory is a restricted theory. The so-called strong constraint, which has its origins in the level matching condition,  restricts the theory to live on a $d$-dimensional subspace of the full $2d$ dimensional doubled space. The strong constraint can be expressed as:
\begin{eqnarray}
\eta_{MN}\pd^M\pd^N\bkt{\cdots}=0,\label{eq:Sconst}
\end{eqnarray}
where `$\cdots$' contains any arbitrary field, parameter or their product. 

The action (\ref{eq:Sdft}) is written in terms of covariant quantities and hence it has a manifest, global $O\bkt{d,d}$ symmetry. This is the $T$-duality symmetry made manifest in the double field theory. Apart from this global symmetry, double field theory has a gauge symmetry which can be interpreted as a symmetry under generalized diffeomorphisms of the doubled space \cite{hohm_large_2013a,naseer_note_2015a}. Generalized diffeomorphisms combine the gauge transformations of $b$-field and the diffeomorphisms  of the metric in an $O\bkt{d,d}$ covariant fashion. Under a generalized coordinate transformation $X^M\rightarrow X^{\prime M}=X^M-\zeta^M$, where $\zeta^M$ is an infinitesimal parameter, the transformation of fields is generated by generalized Lie derivative, i.e.,
\begin{eqnarray}
\delta_{\zeta}\gm_{MN}&=&\gld{\zeta}\gm_{MN}=\zeta^P\pd_P\gm_{MN}+2\bkt{\pd_{(M}\zeta^P-\pd^P\zeta_{(M}}\gm_{N)P},\\
\delta_{\zeta}d&=&\gld{\zeta}d=\zeta^P\pd_Pd-\frac{1}{2}\pd_P\zeta^P\label{eq:gaugeT}
\end{eqnarray}
The algebra of gauge transformations is characterized by the C-bracket \cite{hull_gauge_2009a} i.e.,
\begin{eqnarray}
\sbkt{\delta_{\xi_1},\delta_{\xi_2}}=-\gld{\sbkt{\xi_1,\xi_2}_{\text{C}}},
\end{eqnarray}
where the C-bracket is defined in equation (\ref{eq:defCbra}).
This algebra does not satisfy the Jacobi identity  so generalized diffeomorphisms do not form a Lie group. However, the failure to satisfy the Jacobi identity is of a trivial type and it does not generate a gauge transformation when acting on fields.
\subsection{Frame field formulation}\label{subsec:framefield}
Here we review the frame field formalism for double field theory. Such a formalism was first provided in \cite{siegel_superspace_1993} and its connection with the generalized metric formulation was explained in  \cite{hohm_framelike_2011}. In this formalism, one works with a frame field $E_{A}^{\ M}$. We call the indices $M,N,\cdots$, `curved' indices and $A,B,\cdots,$ `flat' indices. The frame field is subject to a tangent space gauge group. Here, it is convenient to choose the frame field to be a proper element of $O\bkt{d,d}$ as has been done in \cite{geissbuhler_double_2011a}.
\begin{eqnarray}
E_{A}^{\ M}E_{B}^{\ N}\eta_{MN}=\eta_{AB},
\end{eqnarray}
i.e., the $O\bkt{d,d}$ metric with `flat' indices takes the same form as with curved indices and it is used to raise and lower `flat' indices. The generalized metric can then be defined as:
\begin{eqnarray}
\gm_{MN}=E_{M}^{\ A}E_{N}^{\ B}\gm_{AB}.
\end{eqnarray}
 $\gm_{AB}$ is the `flat' generalized metric given by:
\begin{eqnarray}
\gm_{AB}=\begin{pmatrix}
h^{ab}&0\\
0&h_{ab}
\end{pmatrix},
\end{eqnarray}
where $h_{ab}$ is the  $d$-dimensional Minkowski metric, $h_{ab}=\text{diag}\bkt{-1,1,\cdots,1}$ with $h^{ab}$ being its inverse. 

In order to write the action of double field theory in terms of the frame field, we introduce generalized coefficients of anholonomy as follows:
\begin{eqnarray}
\Omega_{ABC}&=&3f_{\sbkt{ABC}},\\
f_{ABC}&=&E_{A}^{\ M}\pd_ME_{B}^{\ N}E_{NC}.
\end{eqnarray}
$f_{ABC}$ is not well behaved under generalized coordinate transformations but its completely anti-symmetric part, $\Omega_{ABC}$, transforms as a scalar. Another scalar object, $\Omega_{A}$, can be built with the help of the dilaton and the frame field as follows:
\begin{eqnarray}
\Omega_{A}=-e^{2d}\pd_{M}\bkt{E_{A}^{\ M}e^{-2d}}.
\end{eqnarray}
In terms of these objects, the Lagrangian density of the double field theory can  be expressed as
\begin{eqnarray}
\mathscr{L}_{\text{DFT}}=e^{-2d}\bkt{\frac{1}{4}\gm^{AB}\Omega_{A}^{\ \ CD}\Omega_{BCD}-\frac{1}{12}\gm^{AB}\gm^{CD}\gm^{EF}\Omega_{ACE}\Omega_{BDF}+\gm^{AB}\Omega_{A}\Omega_{B}}.\label{eq:LDFTFF}
\end{eqnarray}

This concludes our review of double field theory.
\subsection{Space/time split of double field theory}\label{subsec:split}
In this subsection, we re-write the action for double field theory by splitting the full, doubled space-time into temporal and spatial parts explicitly. The basic idea is to write an action for double field theory wherein the time coordinate is not  doubled, i.e., fields are independent of the dual time coordinate and only the spatial coordinates are doubled. Such a re-writing of double field theory action, with $n$ non-compact and $D-n$ compact coordinates has been performed in \cite{hohm_gauge_2013}, which we follow closely here.
We start with a double field theory on $2D$-dimensional doubled space with the generalized metric $\widehat{\gm}_{\hat{M}\hat{N}}$ and the dilaton $\widehat{d}$. The `hatted' index $\hat{M}$ is split as follows\footnote{Note a slight departure from the notation used in the previous subsection. On $2D$-dimensional doubled space we use `hatted' fields and indices. } : 
\begin{eqnarray}
\ ^{\hat{M}}=\bkt{\ _{0},\ ^{0},\ ^M}\ \ \ \ \ \ _{\hat{M}}=\bkt{\ ^{0},\ _{0},\ _M},
\end{eqnarray}
where $M=1,2,\cdots,d$, with $d=D-1$. Coordinates on the $2D$-dimensional manifold can then be expressed as:
\begin{eqnarray}
X^{\hat{M}}=\begin{pmatrix}
\tilde{t}\\
t\\
X^M
\end{pmatrix}.
\end{eqnarray}
The `flat' indices are also split in the similar fashion,
\begin{eqnarray}
\ ^{\hat{A}}=\bkt{\ _{\bar{0}},\ ^{\bar{0}},\ ^A}\ \ \ \ \ \ _{\hat{A}}=\bkt{\ ^{\bar{0}},\ _{\bar{0}},\ _A},
\end{eqnarray}
where we use $\bar{0}$ to differentiate between `flat' and `curved' time. 

With this split, the Lagrangian density for double field theory in the frame field formalism can be written as:
\begin{eqnarray}
\mathscr{L}=e^{-2\widehat{d}}\bkt{\frac{1}{4}\widehat{\gm}^{\hat{A}\hat{B}}\widehat{\Omega}_{\hat{A}}^{\ \ \hat{C}\hat{D}}\widehat{\Omega}_{\hat{B}\hat{C}\hat{D}}-\frac{1}{12}\widehat{\gm}^{\hat{A}\hat{B}}\widehat{\gm}^{\hat{C}\hat{D}}\widehat{\gm}^{\hat{E}\hat{F}}\widehat{\Omega}_{\hat{A}\hat{C}\hat{E}}\widehat{\Omega}_{\hat{B}\hat{D}\hat{F}}+\widehat{\gm}^{\hat{A}\hat{B}}\widehat{\Omega}_{\hat{A}}\widehat{\Omega}_{\hat{B}}},\label{eq:Ldftfull}
\end{eqnarray}
where all the `hatted' objects are proper adaptations of `un-hatted' objects to this split. In particular we have the frame field denoted by $\widehat{E}_{\hat{A}}^{\ \hat{M}}$ and all other objects are defined in terms of it as before, i.e.,
\begin{eqnarray}
\widehat{\gm}_{\hat{M}\hat{N}}= \widehat{E}_{\hat{M}}^{\ \hat{A}} \widehat{E}_{\hat{N}}^{\ \hat{B}} \widehat{\gm}_{\hat{A}\hat{B}},\ \ \ \ \ && \ \ \ \ \ \ \widehat{E}_{\hat{A}}^{\ \hat{M}} \widehat{E}_{\hat{B}}^{\ \hat{N}} \widehat{\eta}_{\hat{M}\hat{N}}=\widehat{\eta}_{\hat{A}\hat{B}},\\
\widehat{\Omega}_{\hat{A}\hat{B}\hat{C}}=3\widehat{E}_{[\hat{A}}^{\ \hat{M}}\pd_{\hat{M}}\widehat{E}_{\hat{B}}^{\ \hat{N}}\widehat{E}_{\hat{N}\hat{C}]},\ \ \ \ \ &&\ \ \ \ \ \ \widehat{\Omega}_{\hat{A}}=-e^{2\widehat{d}}\pd_{\hat{M}}\bkt{\widehat{E}_{\hat{A}}^{\ \hat{M}}e^{-2d}}.
\end{eqnarray}
The generalized metric with the `flat' indices, $\widehat{\gm}_{\hat{A}\hat{B}}$ has the following form:
\begin{eqnarray}
\widehat{\gm}_{\hat{A}\hat{B}}=\begin{pmatrix}
-1 & 0 & 0\\
0 & -1 & 0\\
0&0&\delta_{AB}
\end{pmatrix},\label{eq:genmetFlat}
\end{eqnarray}
where $\delta_{AB}$ is $2d$-dimensional identity matrix. The $O\bkt{D,D}$ invariant metric $\widehat{\eta}_{\hat{M}\hat{N}}$ takes the form:
\begin{eqnarray}
\widehat{\eta}_{\hat{M}\hat{N}}=\begin{pmatrix}
0\ & 1\ & 0\\ 1\ & 0\ & 0 \\ 0\ & 0\ & \eta_{MN}
\end{pmatrix},
\end{eqnarray}
where $\eta_{MN}$ is the usual $O\bkt{d,d}$ invariant metric. Note that the `flat' generalized metric $\widehat{\gm}_{\hat{A}\hat{B}}$ is $O\bkt{d,1}\times O\bkt{d,1}$ invariant.

To proceed, we demand that fields are independent of the dual time coordinate, i.e., $\frac{\pd}{\pd\tilde{t}}\bkt{\cdots}=0$. Following \cite{hohm_gauge_2013}, we now give the frame field $\widehat{E}_{\hat{A}}^{\ \hat{M}}$ in the Lorentz gauge fixed form as follows:
\begin{eqnarray}
\widehat{E}_{\hat{A}}^{\ \hat{M}}&=&\begin{pmatrix}
\widehat{E}^{\bar{0}}_{\ 0}\ \ & \widehat{E}^{\bar{0}0}\ \ & \widehat{E}^{0M}  \\
\widehat{E}_{\bar{0}0}\ \ &\widehat{E}_{\bar{0}}^{\ 0}\ \ & \widehat{E}_{\bar{0}}^{\ M} \\
\widehat{E}_{A0}\ \ &\widehat{E}_{A}^{\ 0}\ \ &\widehat{E}_{A}^{\ M}
\end{pmatrix}
=\begin{pmatrix}
\mathrm{N}\ \ &0\ \ &0  \\ -\frac{1}{2}\mathrm{N}^{-1}\mathcal{N}^{K}\mathcal{N}_{K}\ \ &\mathrm{N}^{-1}\ \ & -\mathrm{N}^{-1}\mathcal{N}^M \\ E_{A}^{\ K}\mathcal{N}_{K}\ \ & 0\ \ & E_{A}^{\ M} 
\end{pmatrix},\label{eq:ffansatz}
\end{eqnarray}
where $E_{A}^{\ M}$ is the frame field for the induced generalized metric, $\gm_{MN}=\bkt{EE^t}_{MN}$, on $2d$ dimensional doubled hyper-surface. $\mathrm
N$  is scalar function of coordinates and $\mathcal{N}^{M}$ is an $O\bkt{d,d}$ covariant vector. Due to obvious similarities with the ADM formalism of general relativity \cite{arnowitt_dynamics_2008}, we identify $\mathrm
{N}$ as the generalized lapse function and $\mathcal{N}^M$ as the generalized shift vector. A short calculation shows that this frame field is indeed a proper $O(D,D)$ element. The generalized metric can be computed explicitly via a straightforward calculation and one obtains,
\begin{eqnarray}
\widehat{\gm}_{\hat{M}\hat{N}}&=&\begin{pmatrix}
\widehat{\gm}^{00}&\widehat{\gm}^{0}_{\ 0}&\widehat{\gm}^0_{\ N}\\
\widehat{\gm}_{0}^{\ 0}&\widehat{\gm}_{00}&\widehat{\gm}_{0N}\\
\widehat{\gm}_{M}^{\ 0}&\widehat{\gm}_{M0}&\widehat{\gm}_{MN}
\end{pmatrix},\\ 
&=&\begin{pmatrix}
-\mathrm{N}^{-2}& \boldsymbol{\alpha} &\mathrm{N}^{-2}\mathcal{N}_N\\
\boldsymbol{\alpha}& -\frac{1}{2}\boldsymbol{\alpha}\mathcal{N}^K\mathcal{N}_K-\mathrm{N}^2+\gm_{PK}\mathcal{N}^P\mathcal{N}^K&-\boldsymbol{\alpha}\mathcal{N}_N+\gm_{NK}\mathcal{N}^K\\
\mathrm{N}^{-2}\mathcal{N}_M&-\boldsymbol{\alpha}\mathcal{N}_M+\gm_{MK}\mathcal{N}^K&\gm_{MN}-\mathrm{N}^{-2}\mathcal{N}_M\mathcal{N}_N
\end{pmatrix}\label{eq:FullgenMet}
\end{eqnarray}
where $\boldsymbol{\alpha}=\frac{1}{2}\mathrm{N}^{-2}\mathcal{N}^K\mathcal{N}_K$. We also re-define the dilaton according to:
\begin{eqnarray}
e^{-2\widehat{d}}=\mathrm{N}e^{-2d}.\label{eq:dilReDef}
\end{eqnarray}
This definition is such that $d$ behaves as a scalar density with respect to the generalized coordinate transformations on $2d$-dimensional doubled hyper-surface.

Let us now evaluate the double field theory action for the frame field given in equation (\ref{eq:ffansatz}). After a straightforward computation, we obtain the following non-zero coefficients of anholonomy:
\begin{eqnarray}
\widehat{\Omega}^{\bar{0}}_{\ \bar{0}C}=\mathrm{N}^{-1}E_{C}^{\ M}\pd_{M}\mathrm{N},\ \ \ \ 
\widehat{\Omega}_{\bar{0}BC}=\mathrm{N}^{-1}E_{MC}\mathcal{D}_{t}E_{B}^{\ M},\\
\widehat{\Omega}_{ABC}=\Omega_{ABC},\ \ \ \ 
\widehat{\Omega}_{\bar{0}}=2\mathrm{N}^{-1}\mathcal{D}_{t}d,\ \ \ \ 
\widehat{\Omega}_{A}=\Omega_A-\mathrm{N}^{-1}E_{A}^{\ M}\pd_M \mathrm{N}
\end{eqnarray}
where $\mathcal{D}_{t}$ is a differential operator defined as 
\begin{eqnarray}
\mathcal{D}_{t}\equiv\pd_t-\gld{\mathcal{N}},
\end{eqnarray}
where $\gld{\mathcal{N}}$ is the generalized Lie derivative with respect to the vector $\mathcal{N}^M$. $\Omega_{ABC}$ and $\Omega_{A}$ corresponds to  coefficients of anholonomy for the frame field $E_{A}^{\ M}$ on $2d$-dimensional doubled hyper-surface. We can now evaluate the action by plugging these coefficients of anholonomy in equation (\ref{eq:Ldftfull}). After some algebra, one finds that:
\begin{dmath}
\mathscr{L}=-\mathrm{N}^{-1}e^{-2d}\bkt{4\bkt{\mathcal{D}_td}^2+\frac{1}{8}\mathcal{D}_t\gm_{MN}\mathcal{D}_t\gm^{MN}+\gm^{MN}\pd_M \mathrm{N}\pd_N \mathrm{N}}+\mathscr{L}_{\text{DFT}}\bkt{d-\frac{1}{2}\log\bkt{\mathrm{N}},\gm_{MN}}. \label{eq:Lcomp1}
\end{dmath}
In the above expression, first two terms provide  `kinetic' terms for the dilaton and the generalized metric. The last term,
 $\mathscr{L}_{\text{DFT}}\bkt{d-\frac{1}{2}\log\bkt{\mathrm{N}},\gm_{MN}}$ can be computed easily by replacing the dilaton $d$ with $d-\frac{1}{2}\log{\mathrm{N}}$ in the expression (\ref{eq:Ldft}). After a short computation one finds that:
 \begin{dmath}
\mathscr{L}_{\text{DFT}}\bkt{d-\frac{1}{2}\log\bkt{\mathrm{N}},\gm_{MN}}-\mathrm{N}^{-1}e^{-2d}\gm^{MN}\pd_M\mathrm{N}\pd_N\mathrm{N}=\mathrm{N}e^{-2d}\mathcal{R}\bkt{d,\gm_{MN}} - b_{1},\label{eq:L2+term}
\end{dmath}
where $\mathcal{R}\bkt{d,\gm_{MN}}$ is given precisely by equation (\ref{eq:GenRic}), and $b_1$ is a total derivative term given by:
\begin{eqnarray}
b_1=-\pd_{M}\bkt{\mathrm{N}e^{-2d}\sbkt{\pd_{N}\gm^{MN}-4\gm^{MN}\pd_Nd}},\label{eq:TotalDT1}
\end{eqnarray}
when included in the action, this will correspond to a boundary term. We ignore this term here and will come back to it in the section \ref{sec:bdyTerms}. We conclude this section by giving the final form of the action  of double field theory on $2d+1$-dimensional space-time (up to a boundary term).
\begin{dmath}
S=\int dtd^{2d}X\ \mathscr{L},\label{eq:act2}
\end{dmath}
where
\begin{dmath}
\mathscr{L}=-\mathrm{N}^{-1}e^{-2d}\bkt{4\bkt{\mathcal{D}_td}^2+\frac{1}{8}\mathcal{D}_t\gm_{MN}\mathcal{D}_t\gm^{MN}}+\mathrm{N}e^{-2d}\mathcal{R}\bkt{d,\gm_{MN}}. \label{eq:Lfinal}
\end{dmath}
\section{Canonical formulation of double field theory}\label{sec:CanForm}
In this section we present the canonical formulation of double field theory starting from Lagrangian density (\ref{eq:Lfinal}). We follow the usual procedure of computing the canonical momenta corresponding to dynamical variables and then perform the Legendre transform to compute the Hamiltonian density. We also derive primary and the secondary constraints arising in the canonical formalism.
\subsection{Canonical momenta and the Hamiltonian}\label{subsec:CanMom}
Dynamical variables in the Lagrangian density (\ref{eq:Lfinal}) are $\bkt{\mathrm{N},\ \mathcal{N}^M,\ d,\ \gm_{MN}}$. We denote their canonically conjugate momenta respectively as $\bkt{\Pi_{\mathrm{N}},\ \Pi^{M},\ \Pi_{d},\ \Pi_{MN}}$. These canonical momenta can be computed easily and one obtains:
\begin{eqnarray}
\Pi_{\mathrm{N}}&=&\frac{\delta\mathscr{L}}{\delta\pd_t\mathrm{N}}=0\label{eq:PriC1},\\
\Pi^{M}&=&\frac{\delta\mathscr{L}}{\delta\pd_t\mathcal{N}_{M}}=0\label{eq:PriC2},\\
\Pi_{d}&=&\frac{\delta\mathscr{L}}{\delta\pd_t d}=-8\mathrm{N}^{-1}e^{-2d}\mathcal{D}_{t}d,\label{eq:CanMomd}\\
\Pi_{MN}&=&\frac{\delta\mathscr{L}}{\delta\pd_t \gm^{MN}}=-\frac{1}{4}\mathrm{N}^{-1}e^{-2d}\mathcal{D}_t\gm_{MN}.\label{eq:CanMomH}
\end{eqnarray}
Note that the canonical momenta corresponding to the lapse function and the shift vector are constrained to be zero. This will lead to further constraints at the level of equations of motions which will be discussed later.  It is now a trivial exercise to compute the Hamiltonian density by performing the Legendre transform. \begin{dmath}
\mathscr{H}=\Pi_{d}\pd_td+\Pi_{MN}\pd_t\gm^{MN}-\mathscr{L},
\end{dmath}
where the time derivatives of fields $d$ and $\gm_{MN}$ are to be written in terms of the canonical momenta (\ref{eq:CanMomd},\ref{eq:CanMomH}). A short computation yields the following expression for the Hamiltonian density
\begin{dmath}
\mathscr{H}=-2\mathrm{N}e^{2d}\bkt{\Pi_{MN}\Pi^{MN}+\frac{1}{32}\Pi_{d}\Pi_{d}}+\Pi_{d}\gld{\mathcal{N}}d +\Pi^{MN}\gld{\mathcal{N}}\mathcal{H}_{MN}-\mathrm{N}e^{-2d}\mathcal{R}\bkt{d,\gm_{MN}} ,\label{eq:HamDens}
\end{dmath}
and the  Hamiltonian can be obtained by integrating over $2d$-dimensional doubled hyper-surface, i.e.,
\begin{dmath}
\mathbf{H}=\int d^{2d}X\ \mathscr{H}.\label{eq:Ham}
\end{dmath}

The action (\ref{eq:act2}) can be written in terms of the Hamiltonian density and canonical variables as follows:
\begin{eqnarray}
S=\int dtd^{2d}X\ \bkt{\Pi_{d}\pd_td+\Pi_{MN}\pd_t\gm^{MN}-\mathscr{H}},
\end{eqnarray}
It will be shown in the next subsection that the Hamiltonian density can be written as:
\begin{eqnarray}
\mathscr{H}=\mathrm{N}\mathscr{B}+\mathcal{N}^M\mathscr{C}_M,
\end{eqnarray}
where $\mathscr{B}$ and $\mathscr{C}_M$ depend on the dilaton $d$, the induced generalized metric $\gm_{MN}$ and their canonical momenta, $\Pi_d$ and $\Pi_{MN}$. This form of the action makes manifest the fact that the lapse function and the shift vector appear only as Lagrange multipliers and are not dynamical fields. Also,  in this formulation of action in terms of the Hamiltonian, equations (\ref{eq:CanMomd}) and (\ref{eq:CanMomH}) are no longer definitions of the canonical momenta but they become equations of motion for $\Pi_d$ and $\Pi_{MN}$.


\subsection{Constraints}\label{subsec:constraints}
In the last subsection we saw that the canonical momenta corresponding to the lapse function and the shift vector vanish. In the language of Dirac\cite{dirac_lectures_2001}, they are called `primary' constraints. Consistency of the theory requires that these primary constraint do not  change under time evolution. In general this consistency condition leads to further constraints on the dynamical fields, known as `secondary constraints'. It may be possible in special cases that this consistency condition does not lead to
any new constraint. This happens when the time derivative of primary constraints vanishes after imposing primary constraints. However, we will see that this is not the case for the primary constraints arising in the canonical formulation of double field theory.  

Consistency of the first primary constraint in equation (\ref{eq:PriC1}) implies that:
\begin{eqnarray}
 \pd_t\Pi_{\mathrm{N}}=0.
 \end{eqnarray}
 Using Hamilton equation of motion we see that this consistency condition implies that:
 \begin{eqnarray}
 \frac{\delta\mathscr{H}}{\delta\mathrm{N}}=0.
 \end{eqnarray}
A straightforward calculation leads to the following secondary constraint:
\begin{eqnarray}
\mathscr{B}\bkt{X}=0,\ \ \ \text{where}\ \ \  \mathscr{B}\bkt{X}\equiv-e^{-2d}\mathcal{R}\bkt{d,\mathcal{H}_{MN}}-2e^{2d}\bkt{\Pi_{MN}\Pi^{MN}+\frac{1}{32}\Pi_{d}\Pi_{d}}, \label{eq:defB}
\end{eqnarray}
where fields on the right hand side of the defining equation of $\mathscr{B}\bkt{X}$ are evaluated at the point $X$ on the $2d$-dimensional doubled hyper-surface. Similarly, the consistency of the second primary constraint in equation (\ref{eq:PriC2}) requires the following to hold.
\begin{eqnarray}
 \frac{\delta\mathscr{H}}{\delta\mathcal{N}^{M}}=0.\label{eq:SecCons2}
\end{eqnarray}
To derive the secondary constraint associated with this, we need to look at the part of the Hamilonian density which involves the generalized shift vector. The shift vector $\mathcal{N}_M$ appears in the Hamiltonian density through the generalized Lie derivative terms:
\begin{eqnarray}
\mathscr{H} = \Pi_{d}\gld{\mathcal{N}}d+\Pi^{MN}\gld{\mathcal{N}}\gm_{MN}+\cdots,\label{eq:const2}
\end{eqnarray}
where we have omitted the terms which do not depend on the shift vector.
 Treating $d$ as a density under generalized diffeomorphisms on the $2d$-dimensional doubled hyper-surface, the term involving the dilaton and the shift vector can be written as:
\begin{dmath}
\Pi_{d}\gld{\mathcal{N}}d=\mathcal{N}^{M}\bkt{\Pi_{d}\pd_{M}d+\frac{1}{2}\pd_{M}\Pi_{d}}-\frac{1}{2}\pd_M\bkt{\mathcal{N}^M\Pi_{d}}.\label{eq:const2Phi}
\end{dmath}
Similarly, using the symmetric nature of $\gm_{MN}$ and $\Pi^{MN}$, the second term in equation (\ref{eq:const2}) can be written as:
\begin{dmath}
\Pi^{MN}\gld{\mathcal{N}}\gm_{MN}=\mathcal{N}^{K}\bkt{\Pi^{MN}\pd_K\gm_{MN}-2\pd_M\bkt{\Pi^{MN}\gm_{NK}-\gm^{MN}\Pi_{NK}}}+2\pd_{M}\bkt{\mathcal{N}^K\sbkt{\Pi^{MN}\gm_{NK}-\gm^{MN}\Pi_{NK}}}.\label{eq:Const2H}
\end{dmath}
So, the shift vector dependent term in the Hamiltonian density takes the following form:
\begin{dmath}
\Pi_{d}\gld{\mathcal{N}}d+\Pi^{MN}\gld{\mathcal{N}}\gm_{MN}=\mathcal{N}^{K}\sbkt{\Pi_{d}\pd_{K}d+\frac{1}{2}\pd_{K}\Pi_{d}+\Pi^{MN}\pd_K\gm_{MN}-2\pd_K\bkt{\Pi^{MN}\gm_{NK}-\gm^{MN}\Pi_{NK}}}+b_2,\label{eq:Const2Combined}
\end{dmath}
where, $b_2$ is a total derivative term which, again, corresponds to a boundary term in the Hamiltonian.
\begin{dmath}
b_2=\frac{1}{4}\pd_M\sbkt{-\mathcal{N}^M\Pi_{d}+4\mathcal{N}^K\bkt{\Pi^{MN}\gm_{NK}-\gm^{MN}\Pi_{NK}}}.\label{eq:TotalDT2}
\end{dmath}
Notice that in the definitons of the canonical momenta given in equations (\ref{eq:CanMomd}) and (\ref{eq:CanMomH}), there is an implicit factor of $e^{-2d}$ which ensures that the total derivative term (\ref{eq:TotalDT2}) transforms as a density. This fact will be important when we include this term in the Hamiltonian to write down a boundary.

Using equations (\ref{eq:Const2Combined}) and (\ref{eq:SecCons2}) and ignoring the total derivative term, it is easy to derive the following costraint:
\begin{eqnarray}
\mathscr{C}_K\bkt{X}&=&0,\ \ \ \ \text{where}\\ 
 \mathscr{C}_K\bkt{X}&\equiv& \Pi^{MN}\pd_K\gm_{MN}+\Pi_{d}\pd_{K}d-2\pd_M\bkt{\Pi^{MN}\gm_{NK}-\gm^{MN}\Pi_{NK}}+\frac{1}{2}\pd_{K}\Pi_{d},\label{eq:Const2Sol}
\end{eqnarray}
and as before, fields and canonical momenta on the right hand side of equation (\ref{eq:Const2Sol}) are to be evaluated at point $X$. These secondary constraints are also required to satisfy the same consistency condition as primary constraint, i.e., they should not change under time evolution. In principle, this can lead to further constraints. However, we will see in the next section that these consistency conditions are trivially satisfied and one does not need to impose any more constraints on the dynamical fields. 
We conclude our discussion here by writing the Hamiltonian in terms of these constraint functions. Using the expression for Hamiltonian density (\ref{eq:HamDens}) and the defining equations for constraints (\ref{eq:defB}) and (\ref{eq:Const2Sol})it is easy to see that up to total derivative terms the Hamiltonian density can be written as:
\begin{eqnarray}
\mathscr{H}=\mathrm{N}\mathscr{B}+\mathcal{N}^M\mathscr{C}_M,
\end{eqnarray}
so that up to boundary terms, the Hamiltonian becomes:
\begin{eqnarray}
\mathbf{H}=\int d^{2d}X\ \bkt{\mathrm{N}\mathscr{B}+\mathcal{N}^M\mathscr{C}_M}.\label{eq:defHam}
\end{eqnarray}
Note that for any solution of the double field theory, these constraints must be satisfied and hence the on-shell value of the Hamiltonian as given here is zero.  However, we will see in section \ref{sec:bdyTerms} that inclusion of boundary terms provide finite non-zero values for the Hamiltonian. 
\section{Algebra of Constraints}\label{sec:algebra}
In the last section we saw how two primary constraints yielded two secondary constraints. The purpose of this section is to demonstrate that the story ends here and secondary constraints arising in double field theory do not lead to any more constraints. We start by introducing the notion of smeared constraints and describing the general method of computation. We argue that the invariance of these constraints under time evolution is equivalent to the on-shell closure of the Poisson bracket of algebra of these constraints. Finally we compute the algebra of constraints explicitly and show that it closes on-shell.

\subsection{Generalities}
Let us start by giving the fundamental Poisson bracket relation between fields and their conjugate momenta:
\begin{eqnarray}
\label{eq:fundPB}
\{\mathcal{H}_{MN}\bkt{X},\Pi^{KL}\bkt{Y}\}&=&\delta^{K}_{(M}\delta^{L}_{N)}\ \delta\bkt{X-Y},\\
\{d\bkt{X},\Pi_{d}\bkt{Y}\}&=&\delta\bkt{X-Y},
\end{eqnarray}
where $\delta\bkt{X-Y}$ is the $2d$-dimensional Dirac delta distribution. The lapse function and the shift vector commute with all the fields and canonical momenta. Any  Poisson bracket involving arbitrary functionals of fields and conjugate momenta can be computed using these fundamental relations. The Poisson bracket involving the Hamiltonian is of particular importance as the time derivative of a functional $\mathscr{F}\bkt{X}$, is given by:
\begin{dmath}
\frac{d}{dt}\mathscr{F}\bkt{X}=\{\mathscr{F}\bkt{X},\mathbf{H}\}.
\end{dmath}
So the time derivative of secondary constraints can then be written as:
\begin{eqnarray}
\frac{d}{dt}\mathscr{B}\bkt{X}&=&\{\mathscr{B},\mathbf{H}\}=\int d^{2d}Y\ \mathrm{N}\bkt{Y}\{\mathscr{B}\bkt{X},\mathscr{B}\bkt{Y}\}+\mathcal{N}^M\{\mathscr{B}\bkt{X},\mathscr{C}_{M}\bkt{Y} \},\label{eq:Tdcons1}\\
\frac{d}{dt}\mathscr{C}_N\bkt{X}&=&\{\mathscr{C}_N,\mathbf{H}\}=\int d^{2d}Y\ \mathrm{N}\bkt{Y}\{\mathscr{C}_N\bkt{X},\mathscr{B}\bkt{Y}\}+\mathcal{N}^M\{\mathscr{C}_N\bkt{X},\mathscr{C}_{M}\bkt{Y} \}\label{eq:Tdcons2}.
\end{eqnarray}
Hence, to show that the secondary constraints are preserved under time evolution, we need to compute Poisson brackets among constraints and demonstrate that they are zero when the constraints themselves are satisfied, i.e., we compute the Poisson brackets in equations (\ref{eq:Tdcons1}) and (\ref{eq:Tdcons2}) first and then impose the constraints.
It is easy to see that Poisson brackets involving `bare' constraints, ($\mathscr{B}\bkt{X},\mathscr{C}\bkt{X}$) involve Dirac deltas and their derivatives. Dirac deltas are distributions which are easy to handle under an integration. Therefor, we introduce the notion of `smeared' constraints as follows:
\begin{eqnarray}
\mathbf{B}\sbkt{\lambda}&\equiv& \int d^{2d}X\ \lambda\bkt{X}\label{eq:defSmear1} \mathscr{B}\bkt{X},\\
\mathbf{C}\sbkt{\xi}&=&\int d^{2d}X\ \xi^{M}\bkt{X}\mathscr{C}_{M}\bkt{X}.\label{eq:defSmear2}
\end{eqnarray}
 In the above definition,  $\lambda\bkt{X}$ is an arbitrary function of coordinates such that the integration on the right hand side of the above defining equations is well defined. $\xi^M\bkt{X}$ is a generalized vector function with same properties. The smeared constraints are now just numbers and requiring $\mathbf{B}\sbkt{\lambda}$ and $\mathbf{C}\sbkt{\xi}$ to vanish for all choices of $\lambda\bkt{X}$ and $\xi^M\bkt{X}$ is equivalent to requiring $\mathscr{B}\bkt{X}$ and $\mathscr{C}_{K}\bkt{X}$ to vanish at all points on the doubled hyper-surface. Using smeared constraint makes calculations more straightforward and better defined. In terms of the smeared constraints, the Hamiltonian takes the following form:
\begin{eqnarray}
\mathbf{H}=\mathbf{B}\sbkt{\mathrm{N}}+\mathbf{C}\sbkt{\mathcal{N}},
\end{eqnarray}
and time derivatives of smeared constraints can be neatly expressed by taking their Poisson bracket with the Hamiltonian as follows.
\begin{eqnarray}
\frac{d}{dt}\mathbf{B}\sbkt{\lambda}&=\{\mathbf{B}\sbkt{\lambda},\mathbf{B}\sbkt{\mathrm{N}}\}+\{\mathbf{B}\sbkt{\lambda},\mathbf{C}\sbkt{\mathcal{N}}\},\\
\frac{d}{dt}\mathbf{C}\sbkt{\xi}&=\{\mathbf{C}\sbkt{\xi},\mathbf{B}\sbkt{\mathrm{N}}\}+\{\mathbf{C}\sbkt{\xi},\mathbf{C}\sbkt{\mathcal{N}}\}.
\end{eqnarray}
Hence, to show that secondary constraints are preserved under time evolution it suffices to demonstrate that the Poisson bracket algebra of the smeared secondary constraints actually closes on smeared secondary constraints.
\subsection{Algebra of constraints}
In this subsection, we compute the Poisson bracket algebra of the smeared constraints explicitly and show that it closes. We will introduce some useful notations as we proceed with our calculations. 
\subsection*{$\bullet\ \ \{\mathbf{C}\sbkt{\xi_1},\mathbf{C}\sbkt{\xi_2}\}$}
Using the definition of the bare constraint $\mathscr{C}_{K}$ from equation (\ref{eq:Const2Sol}) in the smeared constraint (\ref{eq:defSmear2}) and doing an integration by parts, the smeared constraint $\mathbf{C}\sbkt{\xi}$ can be written in the following form:
\begin{eqnarray}
\mathbf{C}\sbkt{\xi}
&=&\int d^{2d}X\ \bkt{\Pi_{d}\gld{\xi}d\ +\ \Pi^{MN}\gld{\xi}\mathcal{H}_{MN}},\label{eq:defSmear22}
\end{eqnarray}

Let us now compute the Poisson bracket between two smeared constraints associated with vector functions $\xi_1^M\bkt{X_1}$ and $\xi_2^M\bkt{X_2}$. Direct computation shows that:
\begin{eqnarray}
\{\mathbf{C}\sbkt{\xi_1},\mathbf{C}\sbkt{\xi_2}\}&=&\ \int d^{2d}X_{1}\ d^{2d}X_{2}\ \bkt{\{\Pi_{d}\bkt{X_1}\ ,\gld{\xi_2}d\bkt{X_2}\}\gld{\xi_1}d\bkt{X_1}\ \Pi_{d}\bkt{X_2} - \bkt{1\leftrightarrow 2}}\nonumber \\
&+&\bkt{\{\Pi^{MN}\bkt{X_1},\gld{\xi_2}\mathcal{H}_{KL}\bkt{X_2}\}\ \gld{\xi_1}\mathcal{H}_{MN}\bkt{X_1}\ \Pi^{KL}\bkt{X_2} - \bkt{1\leftrightarrow 2}}\label{eq:PBCC}.
\end{eqnarray}
Let us focus on the term involving the dilaton. Since, $\gld{\xi}d\bkt{X}=\xi^P\partial_Pd\bkt{X}-\frac{1}{2}\partial_{P}\xi^P$, we deduce that:
\begin{eqnarray}
\{\Pi_{d}\bkt{X_1},\gld{\xi_2}d\bkt{X_2}\}=-\xi_{2}^P\frac{\partial}{\partial X^P_2} \delta\bkt{X_2-X_1},
\end{eqnarray}
We use this relation in equation (\ref{eq:PBCC}). We use integration by parts to remove the derivative acting on the delta function and then integrate over $X_2$. We replace the dummy integration variable $X_1$ with $X$. The first line in equation (\ref{eq:PBCC}) finally becomes:
\begin{eqnarray}
 -\int d^{2d}X \ \bkt{ \ \Pi_{d}\bkt{X}\ \xi_{2}^P \partial_P\gld{\xi_1}d\bkt{X} - \bkt{1\leftrightarrow 2}}.
\end{eqnarray}
Now, a short computation shows that 
\begin{eqnarray}
\bkt{ \xi_{2}^P\partial_P\gld{\xi_1}d\bkt{X} - \bkt{1\leftrightarrow 2}}
&=& \gld{\sbkt{\xi_2,\xi_1}_C}d\bkt{X},
\end{eqnarray}
Hence, we conclude that the first line in equation (\ref{eq:PBCC}) is just given by:
\begin{eqnarray}
\int d^{2d} X\ \Pi_{d}\gld{\sbkt{\xi_1,\xi_2}_C}d\bkt{X}.\label{eq:dterm}
\end{eqnarray}

Now, let us focus on the term involving the generalized metric and the corresponding canonical momentum. Using the expression for the generalized Lie derivative of the generalized metric and the symmetry properties arising from the fact that in equation (\ref{eq:PBCC}), the indices inside the Poisson bracket are contracted with symmetric tensors outside the Poisson bracket, we find that:
\begin{dmath}
\{\Pi^{MN}\bkt{X_1},\gld{\xi_2}\mathcal{H}_{KL}\bkt{X_2}\}=-\bkt{\delta^M_{(K}\delta^{N}_{L)}\xi_{2}^P\frac{\pd}{\pd X_2^P}\delta\bkt{X_2-X_1}+2\delta^M_{(P}\delta^{N}_{L)}\bkt{\frac{\pd}{\pd X_2^K}\xi_2^P-\frac{\pd}{X_{2P}}\xi_{2K}}\delta\bkt{X_2-X_1}}.
\end{dmath}
We use this expression in equation (\ref{eq:PBCC}) and do similar kind of computations as for the dilaton term. After a long but straightforward computation, we see that the term on the second line in equation (\ref{eq:PBCC}) takes the following nice form:
\begin{eqnarray}
\int d^{2d}X\ \Pi^{MN}\gld{\sbkt{\xi_1,\xi_2}_C}\mathcal{H}_{MN},
\end{eqnarray}
combining this with the dilaton term (\ref{eq:dterm}) and comparing the resulting expression with definiton (\ref{eq:defSmear22}) we obtain the result for the Poisson bracket under consideration:
\begin{eqnarray}
\{\mathbf{C}\sbkt{\xi_1},\mathbf{C}\sbkt{\xi_2}\}=\mathbf{C}\sbkt{\sbkt{\xi_1,\xi_2}_C}.
\end{eqnarray}
\subsection*{$\bullet\ \ \{\mathbf{B}\sbkt{\lambda},\mathbf{B}\sbkt{\rho}\}$}
Now we turn to the smeared constraint $\mathbf{B}\sbkt{\lambda}$. After using the definition of $\mathscr{B}\bkt{X}$ in equation (\ref{eq:defSmear1}), one obtains:
\begin{eqnarray}
\mathbf{B}\sbkt{\lambda}
&=&-\int d^{2d}X\ \lambda \bkt{e^{-2d}\mathcal{R}+2e^{2d}\bkt{\Pi_{MN}\Pi^{MN}+\frac{1}{32}\Pi_{d}^2}}.
\end{eqnarray}
To proceed, we introduce a useful piece of notation to denote the dependence of fields and parameters on coordinates. Let $\mathcal{O}$ be a function of coordinates, we define:
\begin{eqnarray}
\mathcal{O}_{i}\equiv \mathcal{O}\bkt{X_i}.\label{eq:notation}
\end{eqnarray}
This notation will prove useful when computing  Poisson brackets in which we have to deal with two integration variables $X_1$ and $X_2$. 
 
 It is easy to see that the Poisson bracket between the constraints $
 \mathbf{B}\sbkt{\lambda}$ and $\mathbf{B}\sbkt{\rho}$ has three non trivial terms and can be written in the following form:
\begin{eqnarray}
\{\mathbf{B}\sbkt{\lambda},\mathbf{B}\sbkt{\rho}\}= F+G+H,
\end{eqnarray}
and the three terms are:
\begin{eqnarray}
F&=&\int d^{2d}X_1d^{2d}X_2\ \lambda_1\rho_2\sbkt{\frac{1}{16}e^{-2\bkt{d_1-d_2}}\{\mathcal{R}_1,\Pi_{d 2}^2\}-\bkt{1\leftrightarrow 2}},\\
G&=&\int d^{2d}X_1d^{2d}X_2\ \lambda_1\rho_2\sbkt{\frac{1}{16}e^{2d_2}\{e^{-2d_1},\Pi_{ d 2}^2\}-\bkt{1\leftrightarrow 2}},\\
H&=&\int d^{2d}X_1d^{2d}X_2\ \lambda_1\rho_2\sbkt{2e^{-2\bkt{d_1-d_2}}\{\mathcal{R}_1,\Pi_{MN2}\Pi_2^{MN}\}-\bkt{1\leftrightarrow 2}}.
\end{eqnarray}
Let us compute these terms one by one now. Using the expression for the generalized curvature scalar and the fundamental Poisson brackets (\ref{eq:fundPB}), one finds that:
\begin{eqnarray}
F=\frac{1}{8}\int d^{2d}X_1d^{2d}X_{2}\ \lambda_{1}\rho_{2}
\left[ e^{-2\bkt{d_1-d_2}}\Pi_{d 2} \left( 4\mathcal{H}_1^{MN}\pd_M^1\pd_N^1-8\mathcal{H}_1^{MN}\pd_M^1d_1\pd_N^1\right.\right.\nonumber \\
 \left.\left.+4\pd_M^1\mathcal{H}_{1}^{MN}\pd^1_N\right) \delta\bkt{X_1-X_2}\ -\ \bkt{1\leftrightarrow 2}\right] .
\end{eqnarray}
Now we use the fact that $\pd_{M}^1\delta\bkt{X_1-X_2}=-\pd_{M}^2\delta\bkt{X_1-X_2}$. After some algebra which involves repeated use of integration by parts and a relabeling of the dummy integration variables we see that:
\begin{eqnarray}
F
&=&\int d^{2d}X \ \Pi_{d}\gld{\chi}d,
  \end{eqnarray}
where $\chi^M$ is defined as follows:
\begin{eqnarray}
\chi^M\equiv \lambda\mathcal{H}^{MN}\pd_N\rho-\rho\mathcal{H}^{MN}\pd_N\lambda.\label{eq:defChi}
\end{eqnarray}

The second term, i.e., $G$ is easy to compute and it vanishes. The key observation is that the commutator $\{e^{-2d_1},\Pi_{d_2}\}$ involved in $G$ is simply proportional to $\delta\bkt{X_1-X_2}$ and due to $(1\leftrightarrow 2)$ anti-symmetry the two terms appearing in $G$ cancel each other.

The third term, $H$, is non trivial. To evaluate it, we will need the following Poisson bracket.
\begin{dmath}
\{\mathcal{R}_1,\Pi_{MN2}\Pi_{2}^{MN}\}=2\Pi_{2}^{MN}\bigg[4\pd^1_M\pd^1_Nd_1-\pd_M^1\pd_N^1-4\pd_M^1d_1\pd^1_Nd_1+4\pd^1_Nd_1\pd_M^1 
	+\frac{1}{8}\left(\pd_{M}^1\mathcal{H}^{KL}_1\pd_{N}^1\mathcal{H}_{MN1}+2\mathcal{H}_1^{KL}\pd_{K}^1\mathcal{H}_{MN1}\pd_L^1  \right)-\frac{1}{2}\bkt{\pd^1_{M}\mathcal{H}_1^{KL}\pd_K^1\mathcal{H}_{NL1}+\mathcal{H}_1^{KL}\pd_{M}^1\mathcal{H}_{NL1}\pd_{K}^1+\mathcal{H}_{MR1}\pd^{R1}\mathcal{H}_{NK1}\pd^{K1}}\bigg]\delta\bkt{X_1-X_2}.\nonumber
\end{dmath}
This relation is obtained by using the fundamental Poisson bracket relations (\ref{eq:fundPB}) and the explicit expression for generalized curvature. 
Now we use this result to evaluate $H$. After some manipulations, which are by now familiar, we obtain
\begin{eqnarray}
H=\int d^{2d}X&&\Pi^{MN}\bigg[4\lambda\pd_M\pd_N\rho+\lambda\mathcal{H}^{KL}\pd_K\mathcal{H}_{MN}\pd_L\rho\nonumber \\ &&-\lambda\frac{1}{2}\bkt{\mathcal{H}^{KL}\pd_M\mathcal{H}_{NL}\pd_{K}\rho+\mathcal{H}_{MR}\pd^R\mathcal{H}_{NK}\pd^K\rho}-\bkt{\rho\leftrightarrow\lambda}\bigg],\label{eq:Hcomp}
\end{eqnarray}
  
  Following the result for $F$, we expect that $H$ would also reduce to a similar expression. In the following we will see that this is indeed the case. Recalling the definition of $\chi^M$, a short computation yields the following.
\begin{eqnarray}
&&\int d^{2d}X\ \Pi^{MN}\gld{\chi}\mathcal{H}_{MN} \nonumber \\&=&\int d^{2d}X\ \Pi_{MN}\bigg[\bkt{\lambda\mathcal{H}^{PL}\pd_L\rho\mathcal{H}_{MN}+2\bkt{\pd_M\bkt{\lambda\mathcal{H}^{KL}\pd_L\rho}-\pd^K\bkt{\lambda\mathcal{H}_{ML}\pd^L\rho}}\mathcal{H}_{NK}}\nonumber \\&&~~~~~~~~~~~~~~~~ -\bkt{\lambda\leftrightarrow\rho}\bigg].\label{eq:HLie}
\end{eqnarray}
Now, by using the identities, $\mathcal{H}^{KL}\mathcal{H}_{NK}=\eta^{L}_{\ N}$ and $\mathcal{H}_{NK}\pd_{M}\mathcal{H}^{KL}=-\mathcal{H}^{KL}\pd_M\mathcal{H}_{NK}$, it is easy to see that up to symmetric terms in $\lambda\leftrightarrow\rho$, we have:
\begin{dmath}
\int d^{2d}X\ \Pi^{MN}\pd_{M}\bkt{\lambda\mathcal{H}^{KL}\pd_L\rho}\mathcal{H}_{NK}=\int d^{2d}X\ \Pi^{MN}\lambda\bkt{-\mathcal{H}^{KL}\pd_L\rho\pd_{M}\mathcal{H}_{NK}+\pd_M\pd_N\rho}\label{eq:HLie1}
\end{dmath}
Similarly, up to symmetric terms in $\lambda\leftrightarrow\rho$ we obtain
\begin{dmath}
\int d^{2d}X\ \Pi^{MN}\mathcal{H}_{NK}\pd^{K}\bkt{\lambda\mathcal{H}_{ML}\pd^L\rho}=\int d^{2d}X\ \Pi^{MN}\lambda\bkt{\mathcal{H}_{NK}\pd^K\mathcal{H}_{ML}\pd^L\rho+\mathcal{H}_{NK}\mathcal{H}_{ML}\pd^K\pd^L\rho}\ \ \ \ \ ,\label{eq:HLie2}
\end{dmath}
using (\ref{eq:HLie1}) and (\ref{eq:HLie2}) in equation (\ref{eq:HLie}) and comparing the resulting expression with that in equation (\ref{eq:Hcomp}) one can see that:
\begin{eqnarray}
H=\int d^{2d}X\ \Pi^{MN}\sbkt{\gld{\chi}\mathcal{H}_{MN}+\sbkt{2\lambda\bkt{\pd_M\pd_N\rho+\mathcal{H}_{ML}\mathcal{H}_{NK}\pd^K\pd^L\rho}-\bkt{\lambda\leftrightarrow\rho}}}
\end{eqnarray}
Combining this with the result for $F$ and $G$ we conclude that:
\begin{eqnarray}
\{\mathbf{B}\sbkt{\lambda},\mathbf{B}\sbkt{\rho}\}=\mathbf{C}\sbkt{\chi}+\int d^{2d}X\ \Pi^{MN}\sbkt{2\lambda\bkt{\pd_M\pd_N\rho+\mathcal{H}_{ML}\mathcal{H}_{NK}\pd^K\pd^L\rho}-\bkt{\lambda\leftrightarrow\rho}}.\label{eq:PBBB1}
\end{eqnarray}

The second term on the right hand side in the above equation is identically zero if we use the value of the $\Pi^{MN}$ as determined in equation (\ref{eq:CanMomH}). To see this, we write the integrand in the second term as:
\begin{eqnarray}
\Delta\bkt{\lambda,\rho}-\Delta\bkt{\rho,\lambda},
\end{eqnarray}
where
\begin{eqnarray}
\Delta\bkt{\lambda,\rho}&\equiv&2\lambda\Pi^{MN}\bkt{\eta_{ML}\eta_{NK}+\mathcal{H}_{ML}\mathcal{H}_{NK}}\pd^K\pd^L\rho,\\
&=&-\frac{1}{2}\mathrm{N}^{-1}e^{-2d}\lambda \mathcal{D}_{t}\mathcal{H}^{MN}\bkt{\eta_{ML}\eta_{NK}+\mathcal{H}_{ML}\mathcal{H}_{NK}}\pd^K\pd^L\rho,\label{eq:Delta}
\end{eqnarray}
where the last equality follows by using the on-shell value of $\Pi_{MN}$ given in equation (\ref{eq:CanMomH}). Now, consider the following:
\begin{eqnarray}
\mathcal{H}^{MN}\eta_{NJ}\mathcal{H}^{JP}=\mathcal{H}^{M}_{\ J}\mathcal{H}^{JP}=\eta^{MP}.\label{eq:DeltaManip1}
\end{eqnarray}
By applying  operator $\mathcal{D}_{t}$ on both sides we get the following:
\begin{eqnarray}
\mathcal{H}_{N}^{\ \ P} \mathcal{D}_{t}\mathcal{H}^{MN}+\mathcal{H}^{M}_{\ \ J}\mathcal{D}_{t}\mathcal{H}^{JP}=0.
\end{eqnarray} 
After multiplying this equation by $\eta_{ML}\gm_{PK}$ and a slight relabeling of indices one finds that:
\begin{eqnarray}
\mathcal{D}_{t}\mathcal{H}^{MN}\bkt{\eta_{ML}\eta_{NK}+\mathcal{H}_{ML}\mathcal{H}_{NK}}=0,\label{eq:identity}
\end{eqnarray}
and using this in equation (\ref{eq:Delta}) we see that $\Delta\bkt{\lambda,\rho}$ vanishes identically and the Poisson bracket in equation (\ref{eq:PBBB1}) becomes:
\begin{eqnarray}
\{\mathbf{B}\sbkt{\lambda},\mathbf{B}\sbkt{\rho}\}=\mathbf{C}\sbkt{\chi}.\label{eq:PBBBFinal}
\end{eqnarray}

\subsection*{$\bullet \ \ \{\mathbf{B}\sbkt{\lambda},\mathbf{C}\sbkt{\xi}\}$}
Let us now turn our attention to the Poisson bracket between $\mathbf{B}$ and $\mathbf{C}$. After a short calculation, we can arrange the five different terms appearing in this Poisson bracket as follows:
\begin{eqnarray}
\{\mathbf{B}\sbkt{\lambda},\mathbf{C}\sbkt{\xi}\}&=& T+U+V+W+Z,\label{eq:PBBC}\\
T&=&-\int d^{2d}X_1d^{2d}X_2\ \lambda_1\{e^{-2d_1}\mathcal{R}_1,\Pi_{d 2}\}\gld{\xi_2}d_2,\label{eq:PBBCS}\\
U&=&-\int d^{2d}X_1d^{2d}X_2\ \lambda_1e^{-2d_1}\{\mathcal{R}_1,\Pi_{MN2}\}\gld{\xi_2}\mathcal{H}^{MN}_2,\label{eq:PBBCT}\\
V&=&-\frac{1}{16}\int d^{2d}X_1d^{2d}X_2\ \lambda_1e^{2d_1}\Pi_{d_2}\{\Pi_{d 1}^2,\gld{\xi_2}d_2\},\label{eq:PBBCU}\\
W&=&-2\int d^{2d}X_1d^{2d}X_2\ \lambda_1\Pi^{KL}_2\{\Pi_{MN1}\Pi^{MN}_1,\gld{\xi_2}\mathcal{H}_{KL2}\},\label{eq:PBBCV}\\
Z&=&-2\int d^{2d}X_1d^{2d}X_2\ \lambda_1 \bkt{\Pi_{MN1}\Pi^{MN}_{1}+\frac{1}{32}\Pi_{d 1}^2}\{e^{2d_1},\Pi_{d 2}\}\gld{\xi_2}d_2,\label{eq:PBBCZ}
\end{eqnarray}
where we have used the notation introduced earlier to denote the dependence of fields and parameters on the coordinates $X_1$ and $X_2$, in particular, $\xi_2$ is to be understood as $\xi\bkt{X_2}$.  

Let us compute these terms now. To compute $T$ and $U$, the most convenient approach is to use the fact that the Poisson bracket of an arbitrary function of the canonical field with the corresponding canonical momentum is just the derivative of that function with respect to the canonical field, i.e.,
\begin{eqnarray}
\{\bkt{\mathbf{f}\sbkt{d}}\bkt{X},\Pi_{d}\bkt{Y}\}=\frac{\delta\ \mathbf{f}\sbkt{d}}{\delta  d}\delta\bkt{X-Y},
\end{eqnarray} 
using this fact, it is easy to obtain the following:
\begin{eqnarray}
T&=&-\int d^{2d}X\ \lambda \bigg[\frac{\delta}{\delta d}e^{-2d}\mathcal{R}\bigg]\gld{\xi}d,\\
U&=&-\int d^{2d}X\ \lambda e^{-2d} \bigg[\frac{\delta}{\delta \mathcal{H}_{MN}}\mathcal{R}\bigg]\gld{\xi}\mathcal{H}_{MN}.
\end{eqnarray}
Since the gauge transformations in double field theory are generated by the generalized Lie derivatives \cite{hohm_generalized_2010a}, we can understand $\gld{\xi}d$ and $\gld{\xi}\mathcal{H}_{MN}$ as the gauge transformations of the dilaton and the generalized metric generated by a gauge parameter $\xi$. Now we combine the two terms and write them as follows:
\begin{eqnarray}
T+U=-\int d^{2d}X\ \lambda \delta_{\xi}\bkt{e^{-2d}\mathcal{R}},
\end{eqnarray}
where $\delta_{\xi}$ denotes a gauge transformation generated by parameter $\xi$. Since $\mathcal{R}$ is a gauge scalar and $e^{-2d}$ transforms as a scalar density, we conclude that \cite{hohm_generalized_2010a}:
\begin{eqnarray}
T+U=-\int d^{2d}X\ \lambda \pd_P\bkt{\xi^Pe^{-2d}\mathcal{R}}=\int d^{2d}X\ e^{-2d}\mathcal{R}\xi^P\pd_P\lambda.\label{eq:PBBCST}
\end{eqnarray}

For the next two terms  $U$ and $V$, we resort to the same kind of techniques which were used in computing the Poisson bracket $\{\mathbf{B}\sbkt{\lambda},\mathbf{B}\sbkt{\rho}\}$, i.e., use of fundamental Poisson bracket relations, integration by parts and properties of delta function. The term (\ref{eq:PBBCU}) is somewhat easier to compute and after some algebra we obtain the following:
\begin{eqnarray}
V&=&-\frac{1}{16}\int d^{2d}X\ \lambda  e^{2d}\bkt{\xi^P\pd_P\bkt{\Pi_{d}^2}+2\Pi_{d}^2\pd_P\xi^P}.\label{eq:PBBCU2}
\end{eqnarray}
After using similar methods but rather tedious algebra, we can compute the term in equation (\ref{eq:PBBCV}).
\begin{eqnarray}
W=-2\int d^{2d}X\ \lambda e^{2d}\sbkt{\xi^{P}\pd_{P}\bkt{\Pi_{MN}\Pi^{MN}}+2\Pi_{MN}\Pi^{MN}\pd_P\xi^P}.\label{eq:PBBCV2}
\end{eqnarray}
The last term, (\ref{eq:PBBCZ}), is the easiest one to compute. After a short computation we obtain the following expression:
\begin{eqnarray}
Z=-2\int d^{2d}X\ \lambda \bkt{\xi^P\pd_Pe^{2d}-e^{2d}\pd_P\xi^P} \bkt{\Pi_{MN}\Pi^{MN}+\frac{1}{32}\Pi_{d}^2},\label{eq:PBBCZ2}
\end{eqnarray}
Now we can combine all five terms. After some algebra, all the terms add up nicely to give:
\begin{eqnarray}
T+U+V+W+Z&=&\int d^{2d}X\  \xi^P\pd_P\lambda \sbkt{e^{-2d}\mathcal{R}+2e^{2d}\bkt{\Pi_{MN}\Pi^{MN}+\frac{1}{32}\Pi_{d}\Pi_{d}}},\\
&=& \mathbf{B}\bkt{-\xi^P\pd_P\lambda},
\end{eqnarray}
hence we find that:
\begin{eqnarray}
\{\mathbf{B}\sbkt{\lambda},\mathbf{C}\sbkt{\xi}\}=\mathbf{B}\sbkt{-\xi^P\pd_P\lambda}.
\end{eqnarray}

So we conclude that the algebra of constraints closes on-shell. In particular, only the closure of the bracket $\{\mathbf{B},\mathbf{B}\}$ requires the use of the on-shell value of the canonical momentum $\Pi^{MN}$. As discussed earlier, the closure of the algebra of smeared constraints under Poisson brackets ensures that the constraints are preserved under time evolution.

\section{Conserved charges and applications}\label{sec:conserved}
In this section we construct conserved charges (similar to the notions of ADM energy, momentum etc in general relativity) in double field theory. To construct these charges we need to add appropriate boundary terms to the double field theory Hamiltonian. To do this we need to understand how can the boundary of a doubled space be characterized. This will be the subject of this section.

We start by discussing boundary terms in general relativity and low energy action of NS-NS string. Then we discuss a generalized version of Stokes' theorem which would enable us to write boundary terms in double field theory.  We explicitly show that the boundary terms of double field theory reduce to boundary terms of general relativity and low energy effective action of string theory upon taking proper limits. Afterwards we write the boundary terms for the double field theory Hamiltonian by including the total derivative terms (\ref{eq:TotalDT1}) and (\ref{eq:TotalDT2}) which were neglected earlier. From the boundary terms of the Hamiltonian, we define the notions of generalized ADM energy and momentum which are conserved quantities in double field theory. Finally we apply our results to compute ADM energies and momenta for some known solutions of double field theory.

{ \it Note Added:} While we were finishing this paper, we became aware of the works \cite{Park_Odd_2015} and \cite{Blair_conserved_2015} which also discuss conserved charges in double field theory.
\subsection{Boundary terms}\label{sec:bdyTerms}
\subsection*{Boundary terms in general relativity}

Here we  briefly review the key aspects of the boundary terms in general relativity, for a detailed discussion see \cite{blau_lectures_2014}.

 The Einstein-Hilbert action for general relativity in $d$ space-time dimensions is given by:
\begin{dmath}
S_{\text{EH}}\sbkt{g_{ij}}=\int d^{d}x\  \sqrt{-g}R \label{eq:EHaction},
\end{dmath}
where, $g$ is the determinant of the space-time metric $g_{ij}$ and $R$ is the Ricci scalar, and latin indices  take values $0,1,\cdots, d-1$ To obtain equations of motions we look at the behavior of this action under a variation of the metric. One obtains  \cite{blau_lectures_2014}:
\begin{eqnarray}
\delta S_{\text{EH}}=\int d^4x\ \sqrt{-g}\bkt{R_{ij}-\frac{1}{2}g_{ij}R}\delta g^{ij}\ +\int d^4x\sqrt{-g}\  \nabla_{i}\bkt{\Delta B}^{i},\label{eq:variSEH}
\end{eqnarray}
where the first term gives vacuum Einstein equations in the absence of the boundary terms while the second term involves a total derivative.  $\bkt{\Delta B}^{i}$ is given by:
\begin{dmath}
\bkt{\Delta B}^{k}=\bkt{g^{ki}g^{jl}-g^{kj}g^{il}}\nabla_{j}\delta g_{il}.\label{eq:DB}
\end{dmath}
If the space-time manifold $\mathcal{M}$ does not have a boundary, i.e., $\pd\mathcal{M}=0$, then the second term is zero and the variational problem is well defined and one obtains the vacuum Einstein equations. However, if the space-time manifold is bounded by a hyper-surface $\pd\mathcal{M}$ (for simplicity we only consider time-like boundary), then the second term can be written as a boundary integral using the Stokes' theorem.

Suppose $n^i$ is the unit normal vector to the boundary $\pd\mathcal{M}$, i.e., $g_{ij}n^in^j=1$. If coordinates on the boundary are given by $y^{\bar{i}}$, where $\bar{i}=0,1,\cdots,d-2$, then the Stokes' theorem relates the bulk integral of a divergence to the boundary integral as follows:
\begin{eqnarray}
\int d^{d}x\ \sqrt{-g}\nabla_iJ^i=\int d^{d-1}y \sqrt{-h}n_{i}J^{i}, \label{eq:stoke}
\end{eqnarray}
where $h$ is the determinant of the induced metric $h_{\bar{i}\bar{j}}$ on the boundary $\pd\mathcal{M}$ given by the push-forward of the bulk metric $g_{ij}$, i.e.,
\begin{eqnarray}
h_{\bar{i}\bar{j}}=\frac{\pd x^i}{\pd y^{\bar{i}}}\frac{\pd x^j}{\pd y^{\bar{j}}}\ g_{ij}. \label{eq:indMet}
\end{eqnarray}
One can use the metric $g_{ij}$ and its inverse to define objects $h_{ij}$ and $h^{ij}$ with indices taking values $0,1,\cdots,d-1$. In particular, $h_{ij}$ takes the following form:
\begin{eqnarray}
h_{ij}=g_{ij}-n_{i}n_{j},\ \ \text{such that}\ \ h_{ij}n^i=0.
\end{eqnarray}
Hence, $h_{ij}$ acts as a projector to the boundary $\pd\mathcal{M}$. If the boundary $\pd\mathcal{M}$ is described by a constraint function i.e, $\mathscr{S}\bkt{x}=\text{constant}$, then the normal vector is just related to the gradient of the function $\mathscr{S}\bkt{x}$ as follows:
\begin{eqnarray}
n^{i}\propto g^{ij}\pd_j\mathscr{S}.
\end{eqnarray} It is straightforward to obtain the properly normalized normal vector to the boundary which is given by:
\begin{eqnarray}
n^{i}=\bkt{\frac{1}{\sqrt{g^{ij}\pd_{i}\mathscr{S}\pd_j\mathscr{S}}}}g^{ij}\pd_{j}\mathscr{S}.\label{eq:NormalOrd}
\end{eqnarray}

Using Stokes' theorem we can write the second term in equation (\ref{eq:variSEH}) as follows: 
\begin{eqnarray}
\int d^{d}x\ \sqrt{-g}\nabla_{i}\bkt{\Delta B}^i=\int d^{d-1}y\ \sqrt{-h} n_{i}\bkt{\Delta B}^{i}.\label{eq:DBbdy}
\end{eqnarray}
A short computation shows that the integrand in (\ref{eq:DBbdy}) can be written as:
\begin{eqnarray}
n_{i}\bkt{\Delta B}^{i}=g^{ij}n^{k}\pd_{i} \delta g_{jk}-g^{jk}n^{i}\pd_{i}\delta g_{jk}=h^{ij}n^{k}\pd_{i} \delta g_{jk}-h^{jk}n^{i}\pd_{i}\delta g_{jk}.\label{eq:DBbdy2}
\end{eqnarray}
In the above equation, the first term vanishes if one imposes the standard Dirichlet boundary condition, i.e., $\delta g_{ij}\Large|_{\Sigma}=0$. The second term depends on the normal derivative of $\delta g_{jk}$ and is not zero in general. So, we conclude that, with Dirichlet boundary conditions, the variation of the Einstein-Hilbert action includes a non-zero boundary term:
\begin{eqnarray}
\delta S_{\text{EH}}=\int d^d x\ \sqrt{-g}\bkt{R_{ij}-\frac{1}{2}g_{ij}R}\delta g^{ij}\ -\int d^{d-1}y\sqrt{-h}h^{jk}n^{i}\pd_{i}\delta g_{jk},\label{eq:variSEH2}
\end{eqnarray}
and we see that the requirement $\frac{\delta S_{\text{EH}}}{\delta g_{ij}}=0$, does not lead to vacuum Einstein equations. The way to resolve this is to add a boundary term to $S_{\text{EH}}$ whose variation cancels the boundary term in equation (\ref{eq:variSEH2}). It turns out that the following boundary term achieves this goal\cite{blau_lectures_2014}.
\begin{eqnarray}
S_{\text{EH-bdy}}&=&\int d^{d-1}y\ \sqrt{-h} n_{i}B^{i},\label{eq:Sbdy1}
\end{eqnarray}
where
\begin{eqnarray}
B^{k}=\bkt{g^{ik}\Gamma^{j}_{ji}-g^{ij}\Gamma^{k}_{ij}},\label{eq:Bexp1}
\end{eqnarray}
and a short computation shows that:
\begin{eqnarray}
n_{i}B^{i}=g^{kj}n^{i}\pd_{i}g_{kj}-g^{ij}n^{k}\pd_{i}g_{kj}=h^{kj}n^{i}\pd_{i}g_{kj}-h^{ij}n^{k}\pd_{i}g_{kj},
\end{eqnarray}
so that the total action can be written as:
\begin{eqnarray}
S_{\text{EH}}+S_{\text{EH-bdy}=}\int d^{d}x \sqrt{-g}R+\int d^{d-1}y\ \sqrt{-h}\bkt{g^{ij}n^k\pd_kg_{ij}-g^{ij}n^k\pd_jg_{ki}}\label{eq:SEH+bdy}.
\end{eqnarray}
It is now an easy exercise to verify that the variation of the boundary term exactly cancels the boundary term in the variation of Einstein-Hilbert action under Dirichlet boundary conditions and one obtains:
\begin{dmath}
\delta\bkt{S_{\text{EH}}+S_{\text{EH-bdy}}}\ =\int d^dx\ \sqrt{-g}\bkt{R_{ij}-\frac{1}{2}g_{ij}R}\delta g^{ij}.
\end{dmath}
This boundary term, however, is non-covariant with respect to both, bulk and boundary coordinate transformations. Evidently, this boundary term is not unique as it is defined only up to terms whose variations vanish for Dirichlet boundary conditions. This allows for an improvement and one can indeed introduce a more geometrically transparent boundary term, known as the {\it Gibbons-Hawking-York} (GHY) term \cite{gibbons1977action,PhysRevLett.28.1082}, given by:
\begin{eqnarray}
S_{\text{GHY}}=2\int d^{d-1}y\ \sqrt{-h} K,\label{eq:SGHY}
\end{eqnarray}
where $K$ is the trace of the second fundamental form and is given by
\begin{eqnarray}
K\equiv g^{ij}\nabla_in_j
\end{eqnarray}
It is easy to verify that the difference between  variations of the two boundary terms with respect to the metric vanishes upon imposing Dirichlet boundary conditions. One finds that:
\begin{eqnarray}
S_{\text{GHY}}-S_{\text{EH-bdy}}&=&\int d^{d-1}y\ \sqrt{-h}\bkt{h^{ij}n^{k}\pd_{i}g_{jk}+2h^{i}_{j}\pd_{i}n^j}
\end{eqnarray}
so the difference of the two terms depends on the tangential derivatives (derivative along the boundary $\pd\mathcal{M}$) of the normal vector and the metric. Under Dirichlet boundary conditions, the metric $g_{ij}$ and the normal vector $n^{i}$ are fixed at the boundary and hence we deduce the equivalence of the two boundary terms  (\ref{eq:Sbdy1}) and (\ref{eq:SGHY}).
\subsection*{Boundary terms in low energy effective action}
The low energy effective action of NS-NS sector of conventional string theory is written as:
\begin{eqnarray}
S_{\text{Eff}}=\int d^{d}x\ \sqrt{-g}e^{-2\phi}\bkt{R+4\bkt{\pd\phi}^2-\frac{1}{12}H^2} ,
\end{eqnarray}
where $R$ is the Ricci scalar, $H_{ijk}$ is the field strength associated with the Kalb-Ramond field $b_{ij}$ and is defined as:
\begin{eqnarray}
H_{ijk}=3\pd_{\left[i\right.}b_{jk\left.\right]},
\end{eqnarray}
and $\phi$ is the dilaton. The `kinetic' terms for dilaton and Kalb-Ramond field involve only first the derivative terms and pose a well defined variational problem. In light of our discussion for the boundary terms in general relativity, the Ricci scalar term needs to be compensated with a boundary term to have well defined variational problem. The correct boundary term is the following:
\begin{eqnarray}
S_{\text{Eff-bdy}}=\int d^{d-1}y\ \sqrt{-h}e^{-2\phi}n_{i}B^{i}, \label{eq:LEEbdy}
\end{eqnarray}
where $n^{i}$ is the unit normal to the boundary and $B^{i}$ is given in equation (\ref{eq:Bexp1}). Thus the total action reads:
\begin{dmath}
S_{\text{Eff}}+S_{\text{Eff-bdy}}=\int d^dx \sqrt{-g}e^{-2\phi}\bkt{R+4\bkt{\pd\phi}^2-\frac{1}{12}H^2}+\int d^{d-1}y\ \sqrt{-h}e^{-2\phi}\bkt{g^{ij}n^k\pd_kg_{ij}-g^{ij}n^k\pd_jg_{ki}}\label{eq:Seff+bdy}
\end{dmath}
\subsection*{Boundary terms in double field theory action}
To discuss boundary terms in the context of the double field theory, we need a generalization of Stokes' theorem for a doubled space. We provide such a generalization in appendix \ref{App:GenStoke}. For details on the nature of boundary and gradient vectors associatew with the boundary of a doubled space, we refer the reader to the appendix. Here we will briefly describe the generalized Stokes' theorem.

Consider a $2d$-dimensional doubled space $\mathcal{M}$ with coordinates $X^{M}$, where $M=1,2,\cdots,2d$. The boundary of the doubled space, denoted by $\pd \mathcal{M}$ is $\bkt{2d-1}$-dimensional and it has coordinates $Y^{\bar{M}}$, where $\bar{M}=1,2,\cdots,2d-1$. The boundary is characterized by a gradient vector $\mathbf{N}^M$ defined as:
\begin{eqnarray}
\mathbf{N}_{M}\equiv \pd_M\mathscr{S}\label{eq:defGrad},
\end{eqnarray}
where the boundary is specified by\begin{eqnarray} 
\mathscr{S}\bkt{X}=\text{constant}.\label{def:S}\end{eqnarray}
This gradient vector has to satisfy following constraints originating from the strong constraint:
\begin{eqnarray}
\mathbf{N}^M\pd_M\bkt{\cdots}=0, \ \ \ \ \ \ \mathbf{N}^M\mathbf{N}_M=0,
\end{eqnarray}
where `$\cdots$' contains any fields, parameters or their arbitrary product. These constraints imply that the function $\mathscr{S}\bkt{X}$, which specify the boundary is also subject to the strong constraint. The generalized Stokes' theorem can now be presented as follows:
\begin{eqnarray}
\int d^{2d}X\ \pd_M\bkt{e^{-2d}\mathcal{J}^M}=\int d^{2d-1}Y\ \left|\frac{\pd X}{\pd X^{\prime}}\right| e^{-2d}\mathbf{N}_M\mathcal{J}^{M},\label{eq:sto1}
\end{eqnarray}
where $\mathcal{J}^M$ is a generalized vector and $d$ is the dilaton. $\left|\frac{\pd X}{\pd X^{\prime}}\right|$ is the absolute value of the determinant of the transformation matrix $\frac{\pd X^N}{\pd X^{\prime M}}$, where $X^{\prime M}$ are the coordinates adapted to to the boundary and defined as:
\begin{eqnarray}
X^{\prime M}\equiv\bkt{Y^{\bar{M}},\mathscr{S}}.\label{eq:defAdap}
\end{eqnarray}
The factor $\left|\frac{\pd X}{\pd X^{\prime}}\right| e^{-2d}$ transforms as a density with respect to the generalized diffeomorphisms on the boundary and it is the push-forward of the factor $e^{-2d}$ which is a density with respect to the diffeomorphisms of the bulk space.

Now we turn our attention towards boundary terms in double field theory action. The Lagrangian for double field theory, as given in (\ref{eq:Ldft}) involves only first derivatives of fields and hence does not require any additional boundary term. However, in terms of the generalized curvature this Lagrangian can be written as:
\begin{eqnarray}
\mathscr{L}_{\text{DFT}}=e^{-2d}\mathcal{R}+\pd_{M}\sbkt{e^{-2d}\bkt{\pd_N\gm^{MN}-4\gm^{MN}\pd_Nd}},
\end{eqnarray}
using the generalized Stokes' theorem, we can now write the action for double field theory including an explicit boundary term as follows:
\begin{eqnarray}
S_{\text{DFT}}=\int d^{2d}X\ e^{-2d}\mathcal{R}+\int d^{2d-1}Y\ \left|\frac{\pd X}{\pd X^{\prime}}\right|e^{-2d}\mathbf{N}_M\bkt{\pd_N\gm^{MN}-4\gm^{MN}\pd_Nd}.\label{eq:dft+bdy}
\end{eqnarray}
We stress here that this construction of boundary term for double field theory is superfluous. One can always express the generalized curvature scalar in terms of $\mathscr{L}_{\text{DFT}}$ and get rid of the explicit boundary term. However, it is instructive to write this boundary term explicitly and investigate if it reduces to boundary terms of general relativity and low energy effective action \ref{eq:Seff+bdy}. In the following, we show that this is indeed the case.

Action for general relativity is obtained by demanding that the Kalb-Ramond field, space time dilaton ($\phi$) and dependence on dual coordinate vanish, i.e.,
\begin{eqnarray}
\tilde{\pd}=0, b_{ij}=0,\phi=0.
\end{eqnarray}
This would imply that the generalized metric and the dilaton become:
\begin{eqnarray}
\gm=\begin{pmatrix}
g^{-1}\ &\  0\\
0\ &\ g 
\end{pmatrix}
\ \  \text{and}\ \ \ \ e^{-2d}=\sqrt{-g}.
\end{eqnarray}
The generalized curvature scalar reduces to the Ricci scalar, i.e., $\mathcal{R}=R$. So, the bulk term in (\ref{eq:dft+bdy}) reduces precisely to the Einstein-Hilbert action. 
Let us now focus on the boundary term. Since the boundary has to be completely along the ordinary direction, the coordinates on the bulk space $\mathcal{M}$ and the boundary $\pd\mathcal{M}$ and the gradient vector $\mathbf{N}_M$ can be written as:
\begin{eqnarray}
X^M=\begin{pmatrix}
\tilde{x}_{i}\\ x^i
\end{pmatrix},\ \ \ Y^{\bar{M}}=\begin{pmatrix}
\tilde{x}_{i}\\ y^{\bar{i}}
\end{pmatrix},\ \ X^{\prime M}=\begin{pmatrix}\tilde{x}_{i}\\x^{\prime i}\end{pmatrix}=\begin{pmatrix}
\tilde{x}_{i}\\y^{\bar{i}}\\\mathscr{S}
\end{pmatrix},\ \ \ \mathbf{N}_M=\begin{pmatrix}0\\\pd_i\mathscr{S}\end{pmatrix}\label{eq:reduction}
\end{eqnarray}
Using these expressions, a straightforward calculation shows that:
\begin{eqnarray}
e^{-2d}\mathbf{N}_M\pd_N\gm^{MN}&=&\sqrt{-g}\pd_{i}\mathscr{S}\pd_{j}g^{ij}=-\sqrt{-g}\bkt{g^{kl}\pd_l\mathscr{S}}g^{ij}\pd_jg_{ki},\label{eq:Intermediate1}\\
4e^{-2d}\gm^{MN}\mathbf{N}_M\pd_Nd&=&-\bkt{g^{kl}\pd_l\mathscr{S}}\sqrt{-g}g^{ij}\pd_kg_{ij}\ .\label{eq:Intermediate2}
\end{eqnarray}
To make connection with the boundary terms of general relativity, we would like to express the determinant of the Jacobian in terms of the induced metric. First of all note that for the case at hand:
\begin{eqnarray}
\left|\frac{\pd X}{\pd X^{\prime}}\right|=\left|\frac{\pd x}{\pd x^{\prime}}\right|.
\end{eqnarray}
Now, under the coordinate transformation $x\to x^{\prime}$, the space time metric transforms as follows:
\begin{eqnarray}
g_{ij}\to  g^{\prime}_{ij}=\frac{\pd x^{ k}}{\pd x^{\prime i}}\frac{\pd x^{ l}}{\pd x^{\prime j}}g_{kl},\ \ \ \ \ g^{ij}\to g^{\prime ij}= \frac{\pd x^{\prime i}}{\pd x^{k }}\frac{\pd x^{\prime j}}{\pd x^l}g^{kl}.
\end{eqnarray}
To proceed, we need the following result. Let $M$ be an arbitrary non-singular matrix, then an elementary result from linear algebra implies that the elements of the inverse matrix can be written in terms of the determinant of the matrix $M$ and the corresponding minor as follows\cite{blau_lectures_2014}:
\begin{eqnarray}
\bkt{M^{-1}}_{ij}=\frac{\left|C_{(ij)}\right|}{\left|M\right|},\label{eq:LAresult}
\end{eqnarray}
where $C_{(ij)}$ is the $(ij)$ minor of matrix $M$. Now, let $M$ be the transformed space-time metric, i.e.,
\begin{eqnarray}
M_{ij}=g^{\prime}_{ij}.
\end{eqnarray}
Let us apply the formula (\ref{eq:LAresult}) for ($i=d,j=d$). It is easy to see then:
\begin{eqnarray}
\bkt{M^{-1}}_{dd}=g^{\prime dd}=g^{ij}\pd_i \mathscr{S}\pd_j \mathscr{S},\ \  \ \text{and}\ \ \left|M\right|=-g\left|\frac{\pd x}{\pd x^{\prime}}\right|^2,\label{eq:lem1}
\end{eqnarray}
and the minor $C_{(dd)}$ is equal to the induced metric as defined in equation (\ref{eq:indMet}).
\begin{eqnarray}
\bkt{C_{(dd)}}_{\bar{i}\bar{j}}=\frac{\pd x^{ k}}{\pd y^{\bar{i} }}\frac{\pd x^{ l}}{\pd y^{\bar{j}}}g_{kl}=h_{\bar{i}\bar{j}},\ \ \ \text{so that}\ \ \left|C_{dd}\right|=-h.\label{eq:lem2}
\end{eqnarray}
Using equation (\ref{eq:lem1}) and (\ref{eq:lem2}) in (\ref{eq:LAresult}) we can write the desired determinant in the following way:
\begin{eqnarray}
\left|\frac{\pd X}{\pd X^{\prime}}\right|=\sqrt{\left|\frac{h}{g \bkt{g^{ij}\pd_i\mathscr{S}\pd_j\mathscr{S}}}\right|}.\label{eq:detRel}
\end{eqnarray}
Using this expression and the equations (\ref{eq:Intermediate1}) and (\ref{eq:Intermediate2}), it is easy to see that the integrand of the boundary term of the double field theory action takes the following form:
\begin{dmath}
\left|\frac{\pd X}{\pd X^{\prime}}\right|e^{-2d}\mathbf{N}_M\bkt{\pd_N\gm^{MN}-\gm^{MN}\pd_Nd}=\sqrt{-h}\bkt{\frac{g^{kl}\pd_k\mathscr{S}}{\sqrt{g^{ij}\pd_i\mathscr{S}\pd_j\mathscr{S}}}}\sbkt{g^{ij}\pd_kg_{ij}-g^{ij}\pd_jg_{ki}},
\end{dmath}
We recognize that the factor $\bkt{\frac{g^{kl}\pd_k\mathscr{S}}{\sqrt{g^{ij}\pd_i\mathscr{S}\pd_j\mathscr{S}}}} $ is just equal to the unit normal vector to the boundary as given in equation (\ref{eq:NormalOrd}).
Putting this all together we see that the double field theory action, upon proper truncation reduces to:
\begin{eqnarray}
S_{\text{DFT}}\Large|_{\tilde{\pd}=0,b=0,\phi=0}=\int d^d\tilde{x}\sbkt{\int d^{d}x \sqrt{-g}R+\int d^{d-1}y\ \sqrt{-h}\bkt{g^{ij}n^k\pd_kg_{ij}-g^{ij}n^k\pd_jg_{ki}}}.
\end{eqnarray}
The integration over the dual coordinates just amounts to an overall multiplicative constant and we obtain the standard Einstein-Hilbert action plus the boundary term for general relativity exactly matching equation (\ref{eq:SEH+bdy}).

Let us now see how the action (\ref{eq:dft+bdy}) reduces to the low energy effective action of string theory including the boundary terms as in (\ref{eq:Seff+bdy}). To do this, we solve the strong constraint by requiring that the fields do not depend on the dual coordinates, i.e., $\tilde{\pd}\bkt{\cdots}$=0. The bulk term of the double field theory action then becomes\cite{zwiebach_double_2012a}:
\begin{eqnarray}
\int d^{2d}X\ e^{-2d}\mathcal{R}&=&\int d^d\tilde{x}d^{d}x\sqrt{-g}e^{-2\phi}\bkt{R+4\bkt{\square\phi- \bkt{\pd\phi}^2}-\frac{1}{12}H^2}.
\end{eqnarray}
By doing integration by parts for the term involving $\square\phi$ and after some algebra, the above expression becomes:
\begin{dmath}
\int d^{2d}X\ e^{-2d}\mathcal{R}=\int d^d\tilde{x}d^{d}x\sqrt{-g}e^{-2\phi}\bkt{R+4 \bkt{\pd\phi}^2-\frac{1}{12}H^2}+ 4\int d^{d}\tilde{x}d^{d-1}y \sqrt{-h}e^{-2\phi}n^i\pd_i\phi,
\end{dmath}
and we see that the bulk term of double field theory action yields the bulk term of low energy action plus a boundary term. Let us now focus on the boundary terms in the action of double field theory. Note that the transformation matrix and the gradient vector are the same as in the previous case for general relativity, given in equation (\ref{eq:reduction}). However there is no restriction on the Kalb-Ramond field or the dilaton.

Let us now compute different terms in the boundary action of double field theory. A short computation shows that:
\begin{eqnarray}
 e^{-2d}\mathbf{N}_M\pd_N\gm^{MN}&=&-\sqrt{-g}e^{-2\phi}\bkt{g^{kl}\pd_l\mathscr{S}}g^{ij}\pd_jg_{ki},\\
4e^{-2d}\gm^{MN}\mathbf{N}_M\pd_Nd&=&\sqrt{-g}e^{-2\phi}\bkt{g^{kl}\pd_l\mathscr{S}}\bkt{-g^{ij}\pd_kg_{ij}+4\pd_k\phi}.
\end{eqnarray}
Using the relation (\ref{eq:detRel}), the integrand of the boundary term in the double field theory takes the following form:
\begin{dmath}
\left|\frac{\pd X}{\pd X^{\prime}}\right|e^{-2d}\mathbf{N}_M\bkt{\pd_N\gm^{MN}-\gm^{MN}\pd_Nd}=\sqrt{-h}e^{-2\phi}\bkt{\frac{g^{kl}\pd_k\mathscr{S}}{\sqrt{g^{ij}\pd_i\mathscr{S}\pd_j\mathscr{S}}}}\sbkt{g^{ij}\pd_kg_{ij}-g^{ij}\pd_jg_{ki}-4\pd_k\phi},
\end{dmath}
By identifying the factor $\bkt{\frac{g^{kl}\pd_k\mathscr{S}}{\sqrt{g^{ij}\pd_i\mathscr{S}\pd_j\mathscr{S}}}}$ with the unit normal vector to the boundary (\ref{eq:NormalOrd}), we finally obtain:
\begin{dmath}
S_{\text{DFT}}\Large|_{\tilde{\pd}=0}=\int d^d\tilde{x}\sbkt{\int d^dx \sqrt{-g}e^{-2\phi}\bkt{R+4\bkt{\pd\phi}^2-\frac{1}{12}H^2}+\int d^{d-1}y\ \sqrt{-h}e^{-2\phi}\bkt{g^{ij}n^k\pd_kg_{ij}-g^{ij}n^k\pd_jg_{ki}}},
\end{dmath}
which matches the low energy effective action for NS-NS string including the boundary terms as given in (\ref{eq:Seff+bdy}).
\subsection{Boundary terms in the Hamiltonian and conserved charges}
We recall that in our discussion on the canonical formulation for double field theory we neglected boundary terms. Now we are in a position to better understand those boundary terms and include them in the Hamiltonian. The full Hamiltonian of the double field theory can be written as:

\begin{eqnarray}
\mathbf{H}_{\text{DFT}}=\mathbf{H}+\mathbf{H}_{\text{bdy}},\label{eq:FullH}
\end{eqnarray}
where, $\mathbf{H}$ is the bulk Hamiltonian as defined in equation (\ref{eq:defHam}) and $\mathbf{H}_{bdy}$ is the boundary term which can be written by including the total derivative terms (\ref{eq:TotalDT1}) and (\ref{eq:TotalDT2}). For the boundary $\pd\mathcal{M}$ with coordinates $Y^{\bar{M}}$ and characterized by the constraint function $\mathscr{S}\bkt{X}=\text{constant}$, this boundary term takes the following form: 
\begin{dmath}
\mathbf{H}_{bdy}=\int d^{2d-1}Y\ \left|\frac{\pd X}{\pd X^{\prime}}\right|\mathbf{N}_L\sbkt{e^{-2d} \mathrm{N}\bkt{4\gm^{LP}\pd_Pd-\pd_{P}\gm^{LP}}\ +\ 2\mathcal{N}^P\Pi^{LR}\gm_{RP}-\frac{1}{4}\mathcal{N}^L\Pi_{d}}\label{eq:Hbdy},
\end{dmath}
where $X^{\prime M}$ are the boundary adapted coordinates. $X^{\prime M}$ and $\mathbf{N}_{M}$ are defined as in equations (\ref{eq:defAdap}) and (\ref{eq:defGrad}) respectively. 

As discussed earlier, the on-shell value of the bulk Hamiltonian is zero due to secondary constraints, we conclude that only the boundary term contributes to the on-shell Hamiltonian, i.e., $\mathbf{H}_{\text{on-shell}}=\mathbf{H}_{bdy}$.

Now we can define notions of conserved charges analogous to ADM energy and momentum. However, due to the strong constraint, we need to exercise some care in identifying the correct expressions for conserved charges. 

First, let us make the notion of asymptotic flatness precise for the case of double field theory. Consider the generalized metric $\widehat{\gm}_{\hat{M}\hat{N}}$ on the full $2D$-dimensional doubled space and suppose that it depends on a function $\mathscr{P}$ of coordinates in such a way that it reduces to the flat metric in the limit $\mathscr{P}\rightarrow \infty$. We denote the flat metric by $\widehat{\delta}_{\hat{M}\hat{N}}$ and it is given by:
\begin{eqnarray}
\widehat{\delta}_{\hat{M}\hat{N}}=\begin{pmatrix} -1 &0&0\\0&-1&0\\0&0&\delta_{MN}\end{pmatrix}.\label{eq:FlatGenMet}
\end{eqnarray}
Following the analogy with  general relativity, na\"{\i}vely one would expect that the notion of conserved energy can be defined by taking $\mathrm{N}=1$ and $\mathcal{N}^M=0$ in the boundary Hamiltonian:
\begin{eqnarray}
\mathbf{E}\stackrel{?}{\equiv} \lim_{\mathscr{P}\to\infty} \int_{\pd\mathcal{M}} d^{2d-1}Y\ \left|\frac{\pd X}{\pd X^{\prime}}\right|e^{-2d} \mathbf{N}_L\bkt{4\gm^{LP}\pd_Pd-\pd_{P}\gm^{LP}}\label{eq:Eadm?}.
\end{eqnarray}
This definition of energy is just the numerical value of the $\mathbf{H}_{\text{DFT}}$ for some asymptotically stationary observer in doubled space-time i.e., $\mathrm{N}=1,\mathcal{N}^M=0$. Since $\mathbf{H}_{\text{DFT}}$ does not have explicit time dependence, the conservation of energy $\mathbf{E}$ holds trivially. However we can not identify this definition of conserved energy with a physical quantity. The reason is the strong constraint. Due to the strong constraint, fields and parameters for a particular solution of double field theory only depend on a $d$-dimensional subspace of the $2d$-dimensional doubled hyper-surface. So, we can decompose the doubled hyper-surface into `allowed' subspace $\mathcal{M}_{1}$ and `not-allowed' subspace $\mathcal{M}_2$ with fields being independent of $\mathcal{M}_2$\footnote{In this paper we use the direct product just to denote that fields depend on coordinates on $\mathcal{M}_1$ and are independent of coordinates on $\mathcal{M}_2$. We do not worry about global issues regarding this direct product.}:
\begin{eqnarray}
\mathcal{M}=\mathcal{M}_{1}\times\mathcal{M}_{2}.\label{eq:pro1}
\end{eqnarray}
It is shown in appendix \ref{App:GenStoke} that the boundary of the doubled hyper-surface is also restricted to be along the `allowed' subspace. If the boundary of $\mathcal{M}$ is described by the constraint function $\mathscr{S}\bkt{X}=\text{constant}$, then $\mathscr{S}\bkt{X}$ is only allowed to depend on the sub-space $\mathcal{M}_1$ so that we can write
\begin{eqnarray}
\pd\mathcal{M}=\pd\mathcal{M}_{1}\times\mathcal{M}_{2}.\label{eq:pro2}
\end{eqnarray}
With these considerations in mind, it is easy to see that the expression for energy given in equation (\ref{eq:Eadm?}) is proportional to an undesirable factor, i.e., the volume of the $d$-dimensional `not-allowed' sub-space. So one can define the conserved energy by eliminating the integration over the `not-allowed' sub-space. However, this definition is still not free from ambiguity. The reason for this is that the strong constraint may not completely specify the `allowed' sub-space. To make the argument more precise, suppose the coordinates of the  $2d$-dimensional hyper-surface are arranged as follows:
\begin{eqnarray}
X^M=\bkt{\tilde{y}_i,\tilde{z}_a,y^i,z^a},\ \ i=1,\cdots,m,\ \ \ a=1,\cdots,n,\ \ \ \text{and}\ d=m+n.
\end{eqnarray}
Now assume that fields depend only on coordinates $y^i$ and are independent of $\tilde{y}_i,\tilde{z}_a,z^a$. Then the strong constraint along $z^a$ and dual $\tilde{z}_a$ direction is trivially satisfied, i.e., $\pd_a\bkt{\cdots}=\tilde{\pd}^a\bkt{\cdots}=0$. Hence we can pick any $n$-dimensional slice of the $2n$-dimensional space spanned by $(\tilde{z}_a,z^a)$. Which one should we pick? Note that all the different choices of the $n$-dimensional slice are related by T-duality. So all the choices are allowed and from the perspective of the normal space-time they correspond to different solutions related by T-duality. However, from the perspective of double field theory, we have a single solution which satisfies strong constraint. The choice of the $d$-dimensional sub-space amounts to choosing a duality frame. 

To make this point more precise, consider a solution of double field theory given by constant generalized metric and dilaton, where the generalized metric on the $2d$-dimensional hyper-surface is given by the following line element:
\begin{eqnarray}
ds^2=\gm_{MN}dX^{M}dX^{N}=\bkt{g_{ij}-b_{ik}g^{kl}b_{lj}}dx^idx^j+g^{ij}d\tilde{x}_{i}d\tilde{x}_{j}-2g^{ik}b_{kj}d\tilde{x}_idx^j.\label{eq:soldft1}
\end{eqnarray}
Let us now choose the half-dimensional sub-space of the double space-time spanned by the coordinates $x$ and identify it with the usual space-time. Then, from the space-time perspective, the solution (\ref{eq:soldft1}) corresponds to a solution with space-time metric given by $g_{ij}$ and the Kalb-Ramond field given by $b_{ij}$. On the other hand, since fields are independent of all the coordinates, we can make a different choice. Let us now choose the sub-space spanned by the coordinates $\tilde{x}$ and identify it with the usual space-time. Then, from the space-time perspective, the double field theory solution of equation (\ref{eq:soldft1}) corresponds to a solution with the space-time metric given by:
\begin{eqnarray}
\bkt{\tilde{g}^{-1}}_{ij}=\bkt{g_{ij}-b_{ik}g^{kl}b_{lj}},
\end{eqnarray}
where the role of upper and lower indices has been exchanged in agreement with $x^i\to \tilde{x}_i$, as explained in \cite{hohm_background_2010a}.
Similarly this solution will correspond to a different Kalb-Ramond field from the space-time perspective,  $\tilde{b}_{ij}$. So, we see that the double field theory solution given by (\ref{eq:soldft1}) can lead to different space-time interpretation upon choosing different half-dimensional sub-spaces of the double space-time. Hence, we obtain different values for conserved charges by choosing different sub-spaces of the doubled space-time. A complete understanding of the physical interpretation of these conserved charges from the double field theory perspective must await a better understanding of the geometry of the double space.

With these subtleties in mind we define the conserved energy for a solution of double field theory by eliminating the integration over a half-dimensional subspace in equation (\ref{eq:Eadm?}) as follows:
\begin{eqnarray}
\mathbf{E}\equiv \lim_{\mathscr{P}\to\infty} \int_{\pd\mathcal{M}_{1}} d^{d-1}Y\ \left|\frac{\pd X}{\pd X^{\prime}}\right|e^{-2d} \mathrm{N}_L\bkt{4\gm^{LP}\pd_Pd-\pd_{P}\gm^{LP}}\label{eq:Eadm},
\end{eqnarray}
It is now an easy exercise to show that the above expression indeed reduces to the ADM energy of general relativity.

The notion of conserved momentum can be defined similarly. We have $2d$ conserved quantities associated with spatial translations on the boundary of the doubled space. These quantities are obtained by setting $\mathrm{N}=0$ and $\mathcal{N}^M=1$ ($M=1,2,\cdots,2d$).
\begin{eqnarray}
\mathbf{P}_M\equiv\lim_{\mathscr{P}\to\infty}
\int_{\pd\mathcal{M}_1} d^{d-1}Y\ \left|\frac{\pd X}{\pd X^{\prime}}\right|\sbkt{2\gm_{MK}\Pi^{KL}\mathbf{N}_L-\frac{1}{4}\mathbf{N}_M\Pi_{d}},\label{eq:Padm}
\end{eqnarray}
The conservation of  momentum is also easy to establish.
\subsection{Applications}
This section is devoted to the application of the results obtained previously, i.e., the formulae for conserved energy and momentum in double field theory, to some specific solutions. In particular we will consider the solutions discussed in \cite{berkeley_strings_2014,berman_branes_2015}.
\subsection*{DFT monopole}
DFT monopole solution was discussed in great detail in \cite{berman_branes_2015} and is inspired by the Kaluza-Klein (KK) monopole solution in general relativity \cite{taub_KK,taub_nut_KK}. KK-monopole is a solution of general relativity in five space-time dimensions with one spatial dimension compactified and a divergence localized in the other three directions. As we will see below, the DFT-monopole has a divergence which is localized in a three dimensional subspace of the full doubled space which has other (not necessarily compact) directions as well. Therefore it is appropriate to interpret this solution as `generalized KK-brane'.

To describe this solution, we split the generalized coordinates $X^{\hat{M}}$ as follows:
\begin{eqnarray}
X^{\hat{M}}=\bkt{t,\tilde{t},X^M}=\bkt{\tilde{t},t,\tilde{x}_a,x^a,\tilde{y}_i,y^i,\tilde{z},z},
\end{eqnarray}
where $a=1,\cdots,6$ and $i=1,2,3$ so that the doubled space is $20$ dimensional, i.e., ($D=10,d=9$). The generalized metric for the DFT monopole is described by the following line element:
\begin{eqnarray}
ds^2&=&\widehat{\gm}_{\hat{M}\hat{N}}dX^{\hat{M}}dX^{\hat{N}},\\
&=&-\bkt{dt^2+d\tilde{t}^2}+\delta_{ab}dx^adx^b+\delta^{ab}d\tilde{x}_ad\tilde{x}_b+f\bkt{\delta_{ij}+\frac{A_iA_j}{f^2}}dy^idy^j\nonumber \\
&&+\frac{1}{f}\delta^{ij}d\tilde{y}_id\tilde{y}_j +f\bkt{1+\frac{A^2}{f^2}}dz^2+\frac{1}{f}d\tilde{z}^2 \nonumber\\
&&+\frac{2}{f}A_{i}\bkt{dy^id\tilde{z}-\delta^{ij}d\tilde{y}_jdz},\label{eq:MonoGM}
\end{eqnarray}
where $f$ is a harmonic function of $y^{i}$ coordinates only given by:
\begin{eqnarray}
f\bkt{r}=1+\frac{h}{r},\ \ \ \ \ r^2=\delta_{ij}y^iy^j.
\end{eqnarray}
And the solution for dilaton is given by:
\begin{eqnarray}
e^{-2\widehat{d}}=f.\label{eq:MonoDilaton}
\end{eqnarray}
 $A_{i}$ is a three dimensional divergence less vector which satisfies:
\begin{eqnarray}
\vec{\nabla}\times\vec{A}=\vec{\nabla}f.
\end{eqnarray}
The precise form of $A_i$ is not important here but it is easy to see that it vanishes in  $r\rightarrow\infty$ limit. To recast this solution in terms of the lapse function and shift vector, it is useful to record the non zero component of the generalized metric in the following way:

\begin{eqnarray}
\widehat{\gm}^{00}=\widehat{\gm}_{00}=-1,\ \widehat{\gm}^{ab}=\delta^{ab},\ \widehat{\gm}_{ab}=\delta_{ab},\ \widehat{\gm}_{zz}=f\bkt{1+\frac{A^2}{f^2}}, \ \widehat{\gm}^{\tilde{z}\tilde{z}}=\frac{1}{f},\nonumber \\
 \widehat{\gm}_{ij}=f\bkt{\delta_{ij}+\frac{A_iA_j}{f^2}},\ \widehat{\gm}^{ij}=\frac{1}{f}\delta^{ij},\ 
\widehat{\gm}_{i}^{\ \tilde{z}}=\widehat{\gm}^{\tilde{z}}_{\ i}=\frac{1}{f}A_i,\widehat{\gm}^{j}_{\ z}=\widehat{\gm}^{\ j}_z=-\frac{1}{f}A_i\delta^{ij},
\end{eqnarray}
where we expect the notation to be self explanatory. By comparing with the form of the generalized metric in equation (\ref{eq:FullgenMet}) it is easy to see that:
\begin{eqnarray}
\mathrm{N}=1,\ \ \ \mathcal{N}^M=0,\ \ \gm_{MN}=\widehat{\gm}_{MN},\ \ e^{-2d}=f\label{eq:mono2}.
\end{eqnarray}
It is easy to see that the canonical momenta associated with $\gm_{MN}$ and $d$ vanish and we deduce that the conserved momentum (\ref{eq:Padm}) for the DFT monopole solution vanishes.

Note that in the limit $r\rightarrow\infty$ the generalized metric reduces to the flat form ,i.e.,
\begin{eqnarray}
\widehat{\gm}_{\hat{M}\hat{N}}\stackrel{r\rightarrow\infty}{\longrightarrow}\widehat{\delta}_{\hat{M}\hat{N}}.
\end{eqnarray}
This limit defines the boundary at which we do integration to compute the conserved energy (\ref{eq:Eadm}), i.e., the function $\mathscr{S}\bkt{X}$ of equation (\ref{def:S}) is given by:
\begin{eqnarray}
\mathscr{S}\bkt{X}=r.
\end{eqnarray} 
We will do the integration at a fixed radius $r$ and then take the limit $r\rightarrow \infty$. The gradient vector takes the following form:
\begin{eqnarray}
\mathbf{N}_M=\bkt{
\tilde{n}^a , n_a , \tilde{n}^i, n_i , \tilde{n}^z ,n_z}=\bkt{0 , 0 , 0, \frac{y_i}{r} , 0 ,0}
.
\end{eqnarray}
A short calculation shows that :
\begin{eqnarray}
4e^{-2d}\mathbf{N}_L\gm^{LP}\pd_Pd-e^{-2d}\mathbf{N}_L\pd_P\gm^{LP}=\frac{h}{fr^2}.
\end{eqnarray}
It is natural to use spherical coordinates instead of the Euclidean $(y^1,y^2,y^3)$ at the boundary. So the surface element at the boundary can be written as:
\begin{eqnarray}
d^{2d-1}Y=d^3\tilde{y}d^5xd^5\tilde{x}dzd\tilde{z}d\theta d\phi
\end{eqnarray}
and $\left|\frac{\pd X}{\pd X^{\prime}}\right|$ then provides the correct Jacobian in transforming from the Euclidean $(y^i)$ to spherical $(\theta,\phi)$ coordinates, i.e.,
\begin{eqnarray}
\left|\frac{\pd X}{\pd X^{\prime}}\right|=r^{2}\sin\bkt{\theta}.
\end{eqnarray}
 Let us now choose a particular solution of the strong constraint. An obvious choice is $\tilde{\pd}\bkt{\cdots}=0$. Using this solution for the strong constraint and the definition of conserved energy in equation (\ref{eq:Eadm}), after a short calculation, one obtains:
\begin{eqnarray}
\mathbf{E}=4\pi h\int d^5xdz=4\pi hV_5\int dz,\label{eq:EadmKK}
\end{eqnarray}
 The conserved energy is proportional to a volume element of which can in principle be infinite. As discussed in \cite{berman_branes_2015}, if one treats $z$ as normal coordinate then the DFT monopole solution corresponds to an infinite array of NS5-branes smeared along $z$ direction. The coefficient $4\pi hV_5$ can then be understood as the energy of a single  NS5-brane. This explains the physical origin of $\int dz$ appearing in the expression for conserved energy. On the other hand if one chooses to treat $\tilde{z}$ as the `normal' coordinate and $z$ as the `winding' coordinate then the DFT monopole solution corresponds to a KK-brane solution which is dual to the NS5-brane with KK-circle along the winding direction $z$.
 From the expression for energy obtained above, one can in fact recover the energy for the conventional (5-dimensional) KK-monopole. To do this we set $V_5=1$ and realize that the $z$ direction is compactified into a circle of radius $2h$ \cite{Misner_flatter_1963}. Then the expression (\ref{eq:EadmKK})  reproduces the ADM energy of standard KK-monopole calculated in \cite{Gross198329} (up to an overall  factor of $\frac{1}{16\pi G}$). 
\subsection*{Localized DFT  monopole}
In the previous solution, the harmonic function is independent of the coordinate $z$ and so its divergence is not localized along the $z$ direction.  When reduced to the usual space-time, this solution has the interpretation of NS-5 brane solution in string theory smeared along the $z$ direction \cite{berman_branes_2015}. A solution which is localized in $z$ direction is obtained if the harmonic function is chosen to be:
\begin{eqnarray}
f\bkt{r,z}=1+\frac{h}{r^2+z^2}.
\end{eqnarray}
With this harmonic function, the analysis for conserved energy and momentum is similar to what was done in previous subsection. In particular, the momentum associated with this solution is still zero. The generalized metric reduces to the flat form in the limit $R\equiv \sqrt{r^2+z^2}\rightarrow \infty$. The function $\mathscr{S}\bkt{X}$ describing the boundary is thus given by:
\begin{eqnarray}
\mathscr{S}\bkt{X}=\sqrt{r^2+z^2}=R.
\end{eqnarray}
We perform integration at a fixed value of $R$ and then take the limit $R\rightarrow \infty$. The gradient vector in this case is given by:
\begin{eqnarray}
\mathbf{N}_M=\bkt{
\tilde{n}^a , n_a , \tilde{n}^i, n_i , \tilde{n}^z ,n_z}=
\sqrt{\frac{1}{r^2+z^2}}\bkt{0,0,0,y_i,0,z}.
\end{eqnarray}
 We choose the solution of the strong constraint as $\tilde{\pd}\bkt{\cdots}=0$. Then a straightforward calculation  yields that:
\begin{eqnarray}
\mathbf{E}=4\pi^2hV_5.
\end{eqnarray}
This is exactly the result one would expect for the `mass' of a localized NS5-brane \cite{Sfetsos_rotating_2000}.  

\subsection*{Generalized pp-wave solution}
Here we discuss the generalized pp-wave solution and obtain associated conserved energy and momentum. We will see that this solution actually carries momentum along the $\tilde{z}$ direction as hinted in \cite{berkeley_strings_2014}. To describe the solution, let us split the generalized coordinates $X^{\hat{M}}$ as follows:
\begin{eqnarray}
X^{\hat{M}}=\bkt{\tilde{t},t,X^{M}}=\bkt{\tilde{t},t,\tilde{z},z,\tilde{y}_i,y^i},
\end{eqnarray}
where $i=1,\cdots,d-1$ and the solution is defined on $2D=2(d+1)$ dimensional double space-time. 
The generalized metric is given by the following line element in doubled space:
\begin{dmath}
ds^2=\widehat{\gm}_{\hat{M}\hat{N}}dX^MdX^N,
=\bkt{f-2}dt^2-fd\tilde{t}^2+\bkt{2-f}dz^2+fd\tilde{z}^2+2\bkt{f-1}\sbkt{dtd\tilde{z}+d\tilde{t}dz}
+\delta_{ij}dy^idy^j+\delta^{ij}d\tilde{y}_id\tilde{y}_j.
\end{dmath}
The components of the generalized metric can be written explicitly as follows:
\begin{eqnarray}
\widehat{\gm}_{00}=f-2,\ \widehat{\gm}^{00}=-f,\ \widehat{\gm}_{zz}=2-f,\ \widehat{\gm}^{zz}=f,\ 
\widehat{\gm}^0_{\ z}=\widehat{\gm}^{\ 0}_z=\widehat{\gm}^{z}_{\ 0}=\widehat{\gm}^{\ z}_{0}=f-1.\label{eq:ppwavegen}
\end{eqnarray}
The function $f$ depends only on $y^i$ coordinates and is given by:
\begin{eqnarray}
f=1+\frac{h}{r^{d-3}},\ \ \ r^2=\delta_{ij}y^iy^j.
\end{eqnarray}
The solution for dilaton is given by:
\begin{eqnarray}
e^{-2\widehat{d}}=\mathrm{N}e^{-2d}=\text{constant},\label{eq:gppdil}
\end{eqnarray}
we choose the constant to be unity here.
By comparing (\ref{eq:ppwavegen}) and (\ref{eq:gppdil}) with equations (\ref{eq:FullgenMet}) and (\ref{eq:dilReDef}) we obtain the following:
\begin{gather}
e^{-2d}=\sqrt{f},\ \ \mathrm{N}=\frac{1}{\sqrt{f}},\ \ 
\mathcal{N}_{\tilde{z}}=\mathcal{N}^z=1-\frac{1}{f}, \\
\gm_{zz}=f\bkt{1-\frac{1}{f}}^2+\bkt{2-f},\ \ \gm^{zz}=f\bkt{1-\frac{1}{f}}^2+f, \ \ \gm_{ij}=\delta_{ij},\ \ \ \gm^{ij}=\delta^{ij}.
\end{gather}

Again we see that the generalized metric reduces to the flat form in the limit $r\rightarrow \infty$ (for $d>3$) so we have $\mathscr{S}\bkt{X}=r$.  We will use the same procedure as above to obtain the conserved energy, the only difference being that the space spanned by coordinates $y^i$ is now $d-1$ dimensional. The gradient vector is given by:
\begin{eqnarray}
\mathbf{N}_M=\bkt{ \tilde{n}^z,n_z, \tilde{n}^i,n_i
}=\frac{1}{r}\bkt{0,0, 0,y_i}.
\end{eqnarray}
It is easy to see that,
\begin{eqnarray}
N_L\pd_P\gm^{LP}=0,\ \ \ 4e^{-2d}N_L\gm^{LP}\pd_Pd=\frac{h\bkt{d-3}}{\sqrt{f}}r^{2-d}.
\end{eqnarray}

To integrate over the boundary we use spherical coordinates. The factor $\left|\frac{\pd X}{\pd X^{\prime}}\right|$ provides the appropriate Jacobian for the transformation from Euclidean coordinates $(y^i)$ to spherical coordinates. So, we have:
\begin{eqnarray}
\left|\frac{\pd X}{\pd X^{\prime}}\right|d^{2d-1}Y=r^{d-2}d\tilde{z}dzd^{d-1}\tilde{y}dS_{d-2},
\end{eqnarray}
where $dS_{d-2}$ is the surface element of $d-2$-dimensional unit sphere. Since the integrand is independent of the angular coordinates, the integration just gives surface area of $(d-2)$ dimensional unit sphere. Dependence on $r$ cancels and one obtains a finite result in $r\rightarrow\infty$ limit. We choose to solve the strong constraint by $\tilde{\pd}\bkt{\cdots}=0$. The final expression for the energy is:
\begin{eqnarray}
\mathbf{E}=2h\bkt{d-3}\frac{\pi^{\frac{d-1}{2}}}{\Gamma\bkt{\frac{d -1}{2}}}\int dz
\end{eqnarray}

Notice that for $d=3$ the energy vanishes. For $d=3$, the generalized metric becomes constant and can be put into flat form via a coordinate transformation and hence it should correspond to zero energy. The coefficient in the above expression for the energy should be understood as the energy density carried by the generalized pp-wave smeared along the $z$ direction.

To compute the conserved momentum, we need to compute  momenta conjugate to $d$ and $\gm_{MN}$.  A short calculation shows that the momentum conjugate to $d$ actually vanishes.
\begin{eqnarray}
\Pi_{d}=0.
\end{eqnarray}
Momentum conjugate to $\gm_{MN}$ has some non-vanishing components which can be computed straightforwardly by using the equation (\ref{eq:CanMomH}). After a short computation one finds that the non zero components of $\Pi_{MN}$ are the following:
\begin{eqnarray}
\Pi_{iz}&=&\Pi_{zi}=\Pi^z_i=\Pi^{\ z}_i=\frac{h\bkt{3-d}r^{2-d}}{4f}\sbkt{f\bkt{1-\frac{1}{f}}^2+2-f}\frac{y_i}{r},\\
\Pi^{iz}&=&\Pi^{i}_{\ z}=\Pi^{zi}=\Pi^{\ i}_z=-\frac{h\bkt{3-d}r^{2-d}}{4f}\frac{y^i}{r}.
\end{eqnarray}
 Let's now compute the following vector which appears as integrand in the boundary integral for the conserved momentum.
\begin{eqnarray}
V_M\equiv \gm_{MK}\Pi^{KL}N_{L}=\gm_{MK}\Pi^{Ki}\ \frac{y_i}{r}=\bkt{\gm_{Mz}\Pi^{zi}+\gm_{M}^{\ z}\Pi_{z}^{\ i}}\frac{y_i}{r}.
\end{eqnarray}
 In $r\rightarrow\infty$ limit, it is easy to see that the only non zero components of of $V_M$ are:
\begin{eqnarray}
V^{\tilde{z}}=V_{z}=-\frac{h\bkt{3-d}r^{2-d}}{4}.
\end{eqnarray}
Using this in the formula for conserved momentum and performing the integration as usual we get  momentum along $z$ and the dual $\tilde{z}$ direction given as:
\begin{eqnarray}
\mathbf{P}_z=\mathbf{P}^{\tilde{z}}=\frac{h\bkt{d-3}}{2}\bkt{\frac{\pi^{\frac{d-1}{2}}}{\Gamma\bkt{\frac{d-1}{2}}}}\int dz
\end{eqnarray}
This implies that the generalized pp-wave solution of double field theory carry equal momentum density along $z$ and the dual $\tilde{z}$ direction just as expected.
\section{Conclusions and outlook}\label{sec:conclusions}
We have given the canonical formulation for double field theory. Starting from a double field theory on $2D$-dimensional doubled space, we split the $2D$-dimensional doubled space-time into temporal and spatial parts explicitly. With this split of the coordinates, the generalized metric on the $2D$-dimensional manifold decomposes into the generalized metric on the doubled spatial hyper-surface, the generalized shift vector and the generalized lapse function. This split also required a re-definition of dilaton. In addition we  assumed that the fields are independent of the dual  time coordinate. Hamiltonian for double field theory can then be computed by following the standard procedure. Primary and secondary constraints in the canonical formalism are derived. It is shown that the Poisson bracket algebra of secondary constraints closes on-shell implying that the constraints do not change under time evolution.

To deal with the boundary terms arising in double field theory, we discussed the nature of the gradient vector to the boundary and gave a generalized version of Stokes' theorem. Appropriate boundary terms are added to the Hamiltonian of double field theory and it is then shown that the these boundary terms provide the on-shell value of Hamiltonian. For asymptotically flat doubled space-times, notions of conserved energy and momentum are constructed and result are applied to some known solutions of double field theory.

Our construction of conserved charges provides a convenient way of computing physical observables associated with a particular solution of double field theory. However these conserved charges are only defined for doubled space-times which are asymptotically flat. An interesting direction for future work is to consider doubled space-times which are not asymptotically flat but admit some other asymptotic symmetries. One can then try to generalize the notions of conserved energy and momenta to conserved charges associated with these asymptotic symmetries. 

The expression for conserved energy obtained here involves canonical momenta and fields appearing explicitly. ADM energy of generalized relativity is written in terms of a purely geometric object, i.e., the boundary integral of the trace of extrinsic curvature. Is same sort of geometric understanding possible for the conserved energy of double field theory? This is a promising direction of further investigation and to move forward one needs to develop a theory of surfaces and embeddings in doubled geometry in complete generality. Progress in any of these directions would help obtain a better understanding of the geometry of double field theory.
 
\appendix

\section{Generalized Stokes' theorem}\label{App:GenStoke}
Our goal here is to write the Stokes' theorem for the case of double field theory as in equation (\ref{eq:sto1}). In particular, we assume that our doubled space $\mathcal{M}$ is $2d$ dimensional and it has a $2d-1$ dimensional boundary $\pd\mathcal{M}$ and we want to write the volume integral of a divergence as a boundary integral in terms of a gradient vector to $\mathbf{N}_M$ which characterizes the boundary\footnote{One can try to formulate the boundary integration by introducing an inner product on the doubled space and defining the notion of a normal vector. Although there is a natural candidate for defining the inner product, i.e., the generalized metric \cite{gualtieri_generalized_2004}, we avoid furnishing the doubled space with extra structure and work with the gradient vector.}. We will show that:
\begin{dmath}
\int d^{2d}X\ \pd_M\bkt{e^{-2d}\ \mathcal{J}^M}\ = \int d^{2d-1}Y\ \left|\frac{\pd X}{\pd X^{\prime}}\right|e^{-2d}\ \mathbf{N}_{M}\mathcal{J}^{M},\label{eq:goal}
\end{dmath}
where, $X^M,\ M=1,\cdots,2d$ are coordinates on $\mathcal{M}$, $Y^{\bar{M}},\ \bar{M}=1,\cdots,2d-1$ are coordinates on $\pd\mathcal{M}$. The boundary is specified by $\mathscr{S}\bkt{X}=\text{constant}$ and $X^{\prime M}$ are the coordinates adapted to the boundary, i.e.,
\begin{eqnarray}
X^{\prime M}\equiv \bkt{Y^{\bar{M}},\mathscr{S}}.
\end{eqnarray}
The gradient vector $\mathbf{N}_{M}$ is defined as:
\begin{eqnarray}
\mathbf{N}_{M}=\pd_M\mathscr{S}.
\end{eqnarray}
 A key element in (\ref{eq:goal}) is the gradient vector $\mathbf{N}_M$ and this will be the focus of our discussion below.

\subsection{Conditions on gradient vector}
To develop the general theory of integration on the boundary of a doubled space, it is important to understand the nature of the gradient vector characterizing the boundary. In particular, we will see that the strong constraint puts some non trivial restrictions on the gradient vector. 
Let us now summarize a set of conditions which a suitable gradient vector $\mathbf{N}^M$ should satisfy.
\begin{itemize}
\item If $\mathcal{J}^M$ is of the type `$\cdots\pd^M\cdots$', i.e., the vector index of $\mathcal{J}^M$ is carried by a derivative, then due to the strong constraint, the bulk integral on the left side of equation (\ref{eq:goal}) vanishes. So, we deduce that the gradient vector must satisfy the following.
\begin{eqnarray}
\mathbf{N}^M\pd_M\bkt{\cdots}=0,
\label{eq:cond1}
\end{eqnarray}
where, `$\cdots$' contains any number of fields or their products.
Since `\textit{boundary of a boundary is zero}', we can do a partial integration in the boundary integral to deduce that:
\begin{eqnarray}
\pd_M\mathbf{N}^M=0.\label{eq:cond2}
\end{eqnarray}
\item Now, consider $\mathcal{J}^M=\mathbf{N}^M$, then due to the conditions (\ref{eq:cond1}) and (\ref{eq:cond2}) the bulk term vanishes and the boundary integrand is proportional to $\mathbf{N}_M\mathbf{N}^M$ and we deduce that the gradient vector should satisfy:
\begin{eqnarray}
\mathbf{N}_M\mathbf{N}^M=0.
\label{eq:cond3}
\end{eqnarray}
This confirms the proposal of \cite{berman_boundary_2011} regarding the gradient vector.
The boundary of a doubled space, in general, can be specified by a constraint like:
\begin{eqnarray}
\mathscr{S}\bkt{X}=\text{constant},\label{eq:bdyConstraint}
\end{eqnarray}
where $\mathscr{S} \bkt{X}$ is some function of the coordinates. The gradient vector is then just given by $ \pd_M\mathscr{S} $.  The above conditions are then just a consequence of the strong constraint satisfied by $\mathscr{S}\bkt{X}$. The strong constraint restricts the fields to depend only a half-dimensional subspace of the doubled space $\mathcal{M}$, say $\mathcal{M}_{1}$. We deduce that the boundary of the doubled space is completely along the the subspace $\mathcal{M}_1$, in the sense of equations (\ref{eq:pro1}) and (\ref{eq:pro2}).
This fact plays important role in defining physical quantities for solutions of double field theory.

\end{itemize}
\subsection{General result}
Let us now turn to the proof of the relation (\ref{eq:goal}). We start by showing that the integral (\ref{eq:goal}) does not change under generalized diffeomorphisms.
 The easiest way to see this is to write the integrand of in (\ref{eq:goal}) as follows:
\begin{eqnarray}
\pd_M\bkt{e^{-2d}\mathcal{J}^M}=e^{-2d}\bkt{\pd_M\mathcal{J}^M-2\mathcal{J}^M\pd_M d},
\end{eqnarray}
Now a short calculation shows that the factor multiplying $e^{-2d}$, actually transforms as a scalar, i.e.,
\begin{eqnarray}
\delta_{\xi}\bkt{\pd_M\mathcal{J}^M-2\mathcal{J}^M\pd_M d}=\xi^P\pd_P\bkt{\pd_M\mathcal{J}^M-2\mathcal{J}^M\pd_M d}.
\end{eqnarray}
By using the fact that $e^{-2d}$ transforms as a scalar density, i.e., $\delta_{\xi}e^{-2d}=\pd_P\bkt{\xi^Pe^{-2d}}$, we deduce that the integrand as a whole transforms like a density, i.e.,
\begin{eqnarray}
\delta_{\xi}\bkt{\pd_M\bkt{e^{-2d}\mathcal{J}^M}}=\pd_P\bkt{\xi^P\pd_M\bkt{e^{-2d}\mathcal{J}^M}}.
\end{eqnarray}Then as discussed in section $2.2$ of \cite{hohm_large_2013a}, under a generalized coordinate transformation, $X\to X^{\prime}$, the integrand transforms as follows:
\begin{eqnarray}
\pd_M^{\prime}\bkt{e^{-2d^{\prime}\bkt{X^{\prime}}}\mathcal{J}^{\prime M}\bkt{X^{\prime}}}=\left|\frac{\pd X}{\pd X^{\prime}}\right| \pd_M\bkt{e^{-2d\bkt{X}}\mathcal{J}^M\bkt{X}},
\end{eqnarray}
and the integration measure $d^{2d}X$ transforms with the opposite factor, i.e.,
\begin{eqnarray}
d^{2d}X^{\prime}=\left|\frac{\pd X^{\prime}}{\pd X}\right|d^{2d}X.
\end{eqnarray}
So, we deduce that the integral (\ref{eq:goal}) is invariant under generalized coordinate transformations and we can write \begin{eqnarray}
I=\int d^{2d}X\ \pd_M\bkt{e^{-2d\bkt{X}}\mathcal{J}^M\bkt{X}}=\int d^{2d}X^{\prime} \pd_M^{\prime}\bkt{e^{-2d^{\prime}\bkt{X^{\prime}}}\mathcal{J}^{\prime M}\bkt{X^{\prime}}}.
\end{eqnarray}
Let us consider the following coordinate transformation so that the transformed coordinates are adapted to the boundary, i.e.,
\begin{eqnarray}
\bkt{X^1,X^2,\cdots,X^{2d}}\rightarrow \bkt{X^{\prime 1},X^{\prime 2},\cdots,X^{\prime 2d}}=\bkt{Y^1,Y^2,\cdots,Y^{2d-1},\mathscr{S}},
\end{eqnarray}
since at the boundary $\mathscr{S}=\text{constant}$, it is easy to do the integration using primed coordinates and one obtains that:
\begin{eqnarray}
I=\int d^{2d-1}X^{\prime}\ e^{-2d^{\prime}}\delta_{\ M}^{2d}\mathcal{J}^{\prime M},\label{eq:integ2}
\end{eqnarray}
where the integration measure on the boundary is given by $d^{2d-1}X^{\prime}=d^{2d-1}Y$. We see that the gradient vector  in the primed coordinates ($\mathbf{N}^'_M\bkt{X'}$) is just given by:
\begin{eqnarray}
\mathbf{N}^{\prime}_M\bkt{X^{\prime}}=\delta_{\ M}^{2d}.
\end{eqnarray}
From this we can find the gradient vector in the un-primed coordinates and it is given by\footnote{This is equivalent to the transformation postulated in \cite{hohm_large_2013a} via the matrix $\mathcal{F}\bkt{X^{\prime},X}$, i.e., $\mathbf{N}_{M}\bkt{X}=\mathcal{F}_{M}^{\ N}\bkt{X,X^{\prime}}\mathbf{N}^{\prime}_N\bkt{X^{\prime}}$.}:
\begin{eqnarray}
\mathbf{N}_M\bkt{X}=\frac{\pd X^{\prime P}}{\pd X^M}\mathbf{N}^{\prime}_{P}\bkt{X^{\prime}}= \frac{\pd \mathscr{S}}{\pd X^M}.
\end{eqnarray}
Now by using the fact that $\mathbf{N}^{\prime}_M\mathcal{J}^{\prime M}=\mathbf{N}_M\mathcal{J}^M$, the integral (\ref{eq:integ2}) can be written as:
\begin{eqnarray}
I=\int d^{2d-1}X^{\prime}\ e^{-2d^{\prime}}\mathbf{N}_M\mathcal{J}^M,\label{eq:integ3}
\end{eqnarray}
Now, we use the fact that
\begin{eqnarray}
e^{-2d^{\prime}\bkt{X^{\prime}}}=\left|\frac{\pd X}{\pd X^{\prime}}\right|e^{-2d\bkt{X}}.
\end{eqnarray}
Putting this all together we see that the integral (\ref{eq:integ3}) can be written as:
\begin{eqnarray}
I=\int d^{2d-1}Y\ e^{-2d}\mathbf{N}_M\mathcal{J}^{M}\left|\frac{\pd X}{\pd X^{\prime}}\right|.\label{eq:genStokFinal}
\end{eqnarray}
This is how volume integral of a divergence is related to the boundary integral.
			\acknowledgments
			I would like to thank Olaf Hohm for suggesting this project and for numerous discussions and very insightful comments throughout the course of this work. I would also like to acknowledge several useful comments and suggestions by Barton Zwiebach. This work is supported by the U.S. Department of Energy under grant Contract Number DE-SC00012567.
			\bibliography{library} 
    
			\bibliographystyle{JHEP}
	\end{document}